\documentclass[twocolumn, numberedappendix,appendixfloats]{openjournal}
\usepackage{lineno}

\usepackage{lipsum}

\usepackage{xcolor}
\usepackage{textgreek}
\usepackage[utf8]{inputenc}
\usepackage[english]{babel}
\usepackage{graphicx}
\usepackage{tabularx}
\usepackage{booktabs} 
\usepackage{amsmath}  \usepackage{makecell}
\newcommand{\ve}[2]{$#1\,{\scriptstyle\pm#2}$}   %
\usepackage{hyperref}
\hypersetup{
    unicode, 
    colorlinks=true,
    linkcolor=linkcolor,
    citecolor=linkcolor,
    filecolor=linkcolor,
    urlcolor=linkcolor,
}
\usepackage{color,colortbl}
\definecolor{linkcolor}{rgb}{0.0,0.3,0.5}
\usepackage{tensind}
\tensordelimiter{?}
\DeclareGraphicsExtensions{.bmp,.png,.jpg,.pdf}
\usepackage{verbatim}
\usepackage[normalem]{ulem}
\usepackage{orcidlink}
\usepackage{soul}
\usepackage{amsmath}

\graphicspath{ {./figs/} }
\renewcommand{\andname}{for the}

\begin{document}
\title{Extending SOMPZ beyond Spectroscopic Coverage for Roman Weak Lensing} 

\author{Yufei Zhen\orcidlink{0009-0006-1726-0150}}
\affiliation{Data Science Institute, The University of Chicago, Chicago, IL 60615, USA}

\author{Chun-Hao To\orcidlink{0000-0001-7836-2261}}
\affiliation{Astronomy and Astrophysics Department, The University of Chicago, Chicago, IL 60637, USA}
\affiliation{Kavli Institute for Cosmological Physics, The University of Chicago, Chicago, IL 60637, USA}
\affiliation{NSF-Simons AI Institute for the Sky (SkAI), 172 E. Chestnut St., Chicago, IL 60611, USA}

\author{Boyan Yin\orcidlink{0009-0006-5604-9980}}
\affiliation{Department of Physics, Duke University, Durham, NC 27708, USA}

\author{Junzhou Zhang\orcidlink{0009-0000-3461-4041}}
\affiliation{Department of Physics, The University of Chicago, Chicago, IL 60637, USA}

\author{Chihway Chang\orcidlink{0000-0002-7887-0896}}
\affiliation{Astronomy and Astrophysics Department, The University of Chicago, Chicago, IL 60637, USA}
\affiliation{Kavli Institute for Cosmological Physics, The University of Chicago, Chicago, IL 60637, USA}
\affiliation{NSF-Simons AI Institute for the Sky (SkAI), 172 E. Chestnut St., Chicago, IL 60611, USA}

\author{Finian Ashmead\orcidlink{0000-0001-9416-5874}}
\affiliation{Department of Physics and Astronomy, University of Pittsburgh, Pittsburgh, PA 15260, USA}
\affiliation{Pittsburgh Particle Physics, Astrophysics, and Cosmology Center (PITT PACC), University of Pittsburgh, Pittsburgh, PA 15260, USA}

\author{Brett H.~Andrews\orcidlink{0000-0001-8085-5890}}
\affiliation{Department of Physics and Astronomy, University of Pittsburgh, Pittsburgh, PA 15260, USA}
\affiliation{Pittsburgh Particle Physics, Astrophysics, and Cosmology Center (PITT PACC), University of Pittsburgh, Pittsburgh, PA 15260, USA}

\author{Daniel C. Masters\orcidlink{0000-0001-5382-6138}}
\affiliation{Caltech/IPAC, 1200 E. California Blvd. Pasadena, CA 91125, USA}

\author{Jeffrey A. Newman\orcidlink{0000-0001-8684-2222}}
\affiliation{Department of Physics and Astronomy, University of Pittsburgh, Pittsburgh, PA 15260, USA}
\affiliation{Pittsburgh Particle Physics, Astrophysics, and Cosmology Center (PITT PACC), University of Pittsburgh, Pittsburgh, PA 15260, USA}

\author{Tianqing Zhang}
\affiliation{Department of Physics and Astronomy, University of Pittsburgh, Pittsburgh, PA 15260, USA}
\affiliation{Kavli Institute of Particle Astrophysics and Cosmology, Stanford, CA, 94305, USA}
\affiliation{SLAC National Accelerator Laboratory, Menlo Park, CA 94025, USA}

\author{Roman HLIS Cosmology Project Infrastructure Team}
\noaffiliation

\begin{abstract}

Precision weak lensing cosmology requires accurate galaxy redshift distributions. However, the spectroscopic samples used to characterize these distributions do not fully span the magnitude--color--redshift space of weak lensing source galaxies, necessitating extrapolation. This challenge will be critical at the depth of the Roman Space Telescope’s High Latitude Imaging Survey (Roman HLIS). Using simulated LSST Year 4 and Roman HLIS photometry from the \texttt{Cardinal} simulations, we quantify how spectroscopic incompleteness biases photometric-redshift (photo-$z$) calibration and cosmological constraints. We find that the standard Self-Organizing Map $p(z)$ (SOMPZ) method produces substantial tomographic bin overlaps and biased source redshift distributions. To address these limitations, we develop Roman-SOMPZ, which uses dedicated individual-galaxy photo-$z$ estimators to supply redshift references for tomographic binning and redshift estimation. For the latter, we introduce UMAP-flow-$z$, an estimator that uses a UMAP embedding of deep-field galaxy colors to infer redshifts for phenotypes lacking spectroscopic representation. Compared with the standard pipeline, Roman-SOMPZ produces better-separated tomographic bins and lower absolute mean-redshift biases in eight of nine bins. Propagating the estimated redshift distributions into cosmological inference, we find that existing spectroscopy is adequate for early-stage weak lensing analyses in the Roman deep fields. Conversely, full medium-tier cosmology will require further improvements in redshift characterization.
\end{abstract}


\maketitle

\section{Introduction}
\label{sec:intro}

Weak gravitational lensing provides a sensitive probe of the growth of cosmic structure and the geometry of the universe \citep[see e.g.,][for a review]{Mandelbaum2018}. Extracting cosmological information from weak lensing signals requires accurate redshift estimates of the galaxies involved, because the lensing efficiency depends on the relative distances between the observer, lenses, and sources. Characterizing these redshift distributions becomes more difficult as current imaging surveys reach fainter and more distant galaxy populations \citep{photoz-review}. 

The Nancy Grace Roman Space Telescope's High-Latitude Imaging Survey \citep[HLIS;][]{HLIS} exemplifies this challenge. In the adopted medium-tier configuration, Roman HLIS is expected to achieve an unprecedented weak lensing source density \citep{ROTAC}, which will translate into precise cosmological constraints \citep{Eifler21}. The survey depth that enables this statistical power also places a substantial fraction of the source galaxies in faint, high-redshift regimes, for which representative spectroscopic samples are difficult to obtain. Thus, for most Roman weak lensing source galaxies, redshift information will be inferred from broadband photometry rather than spectroscopy. Following \cite{photoz-review}, we distinguish between photometric-redshift (photo-$z$) \emph{performance}, which describes the precision of redshift estimates for individual galaxies, and photo-$z$ \emph{characterization}, which concerns recovering ensemble redshift distributions in tomographic bins. This work focuses on the latter, as weak lensing cosmology primarily requires accurate knowledge of the source galaxy redshift density function $n(z)$ in each tomographic bin. 

Self-Organizing Map $p(z)$ \citep[SOMPZ;][]{Masters15, SOMPZ, KIDSsompz, SOM-DES_Y6, Euclidsompz} is a photo-$z$ characterization framework that has been extensively applied to the Dark Energy Survey (DES) \citep{SOMPZ-DES_Y3,2024MNRAS.527.2010G, SOMPZ-DES_Y6, 2026PhRvD.114b3520G}. SOMPZ leverages the ``wedding cake'' strategy common to modern imaging surveys, which combines high-quality photometry from small-area deep fields with the larger, shallower wide-field sample used for cosmology. 
A reference sample with high-confidence redshifts is used to calibrate the redshift distributions of the photometric samples. The deep-field photometry helps distinguish galaxy populations that are degenerate in the wide-field measurements, acting as the bridge through which SOMPZ propagates redshift information to the source sample in the wide field. The incomplete spectroscopic coverage of the faint, high-redshift galaxy populations probed by Roman HLIS will complicate the application of SOMPZ. 
We use simulations to quantify how spectroscopic incompleteness affects SOMPZ redshift characterization and develop an extended framework of 
that used by DES (hereafter ``DES-SOMPZ") in anticipation of the Roman HLIS data.

Incomplete spectroscopic coverage already limits existing analyses. \cite{SOMPZ-DES_Y6} identifies the lack of faint, representative spectra as the dominant contributor to redshift-calibration uncertainties of the DES Year Six (DES~Y6) weak lensing source sample. Roman's greater depth and near-infrared coverage that probe fainter and higher-redshift galaxies make this challenge more severe. In the Roman-like \texttt{Cardinal} \citep{Cardinal} catalogs studied here, under DES-SOMPZ, spectroscopic redshifts do not provide a direct calibration reference for regions in photometric space that contain $26.6\%$ of the deep-field galaxies, compared to $5.7\%$ in DES~Y6. It also leaves $7.2\%$ of the wide-field (i.e., Roman medium-tier) sample without tomographic bin assignments, compared to $0\%$ in DES~Y6. These uncalibrated populations preferentially contain faint, high-redshift galaxies and have two primary effects on DES-SOMPZ. When assigning galaxies to tomographic bins, discarding unassigned wide-field galaxy phenotypes reduces the number of usable weak lensing sources. During $n(z)$ estimation, missing redshift information for deep-field phenotypes leads to biased ensemble distributions. A Roman-adapted SOMPZ pipeline must address both effects. 

Training sample augmentation offers one possible solution to address incomplete redshift-reference samples. For example, \cite{Moskowitz2024} and \cite{2026MNRAS.547f2226Z} augment spec-$z$ training samples with simulated galaxies in color--magnitude--redshift regions underrepresented by spectroscopy and demonstrate improvements in individual-galaxy photo-$z$ performance. The missing galaxy population could also be estimated by generative forward models. The \texttt{pop-cosmos} \citep{pop-cosmos, pop-cosmos1} framework, for example, models a population-level distribution of galaxy redshifts and physical properties from deep multiband photometry. Those properties are connected to observed photometry using stellar population synthesis and survey data models. 

Our approach shares this motivation, but differs in both the augmented objects and the target statistics. Rather than adding mock galaxy populations, we supplement the redshift information for the observed deep fields using individual-galaxy photo-$z$ estimators. 
Because these galaxies are already present in the catalog, we do not need a model to reproduce their abundance or photometric properties. Neither does it require an external simulated population whose colors are susceptible to systematic biases. The inferred redshifts nevertheless depend on extrapolation from the incomplete spec-$z$ sample and carry uncertainties associated with the chosen estimators.

We introduce two extensions to the DES-SOMPZ framework to mitigate spec-$z$ incompleteness, forming the Roman-SOMPZ framework. The first uses photo-$z$ point estimates to assign wide-field populations not covered by spectroscopy to tomographic bins and improve binning. The second augments the redshift reference for poorly calibrated deep-field phenotypes, reducing biases in the $n(z)$ estimates. These additional individual-galaxy redshift estimators are trained using the spec-$z$'s and deep-field photometric features, and their performances are evaluated against the simulation-truth redshifts of the target galaxies. We compare the redshift characterization of \texttt{Cardinal} source galaxies from Roman-SOMPZ, DES-SOMPZ, and an idealized case assuming representative spec-$z$ coverage of the deep fields. 

The remainder of this paper is organized as follows. Section~\ref{sec:data} describes the simulated \texttt{Cardinal} catalogs and the spectroscopic selection used in this study. Section~\ref{sec:sompz} reviews the DES-SOMPZ formalism and diagnoses its failure modes under incomplete spectroscopic coverage. Section~\ref{sec:roman_sompz} presents the Roman-SOMPZ extensions for improved tomographic binning and training sample augmentation, and Section~\ref{sec:cosmo} assesses how its estimated $n(z)$ affects cosmological constraints. Section~\ref{sec:conclusion} summarizes our conclusions.

\section{Simulations}
\label{sec:data}

In this work, we use the \texttt{Cardinal} mock catalogs \citep{Cardinal}, a one-quarter-sky synthetic galaxy sample developed to support precision cosmological analyses with current and next-generation large surveys. 
The \texttt{Cardinal} catalogs are populated up to redshift $z=2.3$ to $r$-band depth $r=27$. Their calibrated luminosities, colors, and color-dependent clustering reproduce the properties of observed galaxy populations. The catalogs are generated using the \texttt{ADDGALS} algorithm \citep{ADDGALS} and include realistic galaxy multiband fluxes and photometric errors. 

We adopt the wide and deep \texttt{Cardinal} configurations designed to reproduce the reference Roman High Latitude Wide Area Survey (HLWAS) scenarios \citep{ROTAC}. The wide configuration approximates the Roman Medium Tier Core Community Survey, covering $\sim2000~\mathrm{deg}^2$ in the Y106, J129, and H158 near-infrared bands. We assume a three-dither, two-pass strategy with $107~\mathrm{s}$ exposures. For exponential-profile sources with a scale length of $0.3''$, this strategy yields a $5\sigma$ limiting magnitude of $m_5=25.4$ in each band. Within this footprint, we simulate two $9.62~\mathrm{deg}^2$ fields representing the HLWAS deep tier. The deep configuration assumes a three-dither, five-pass strategy with $294~\mathrm{s}$ exposures and has a corresponding depth of $m_5=26.7$. 

\begin{center}
    \begin{figure}[!t]
        \centering
    	\includegraphics[width=\linewidth]{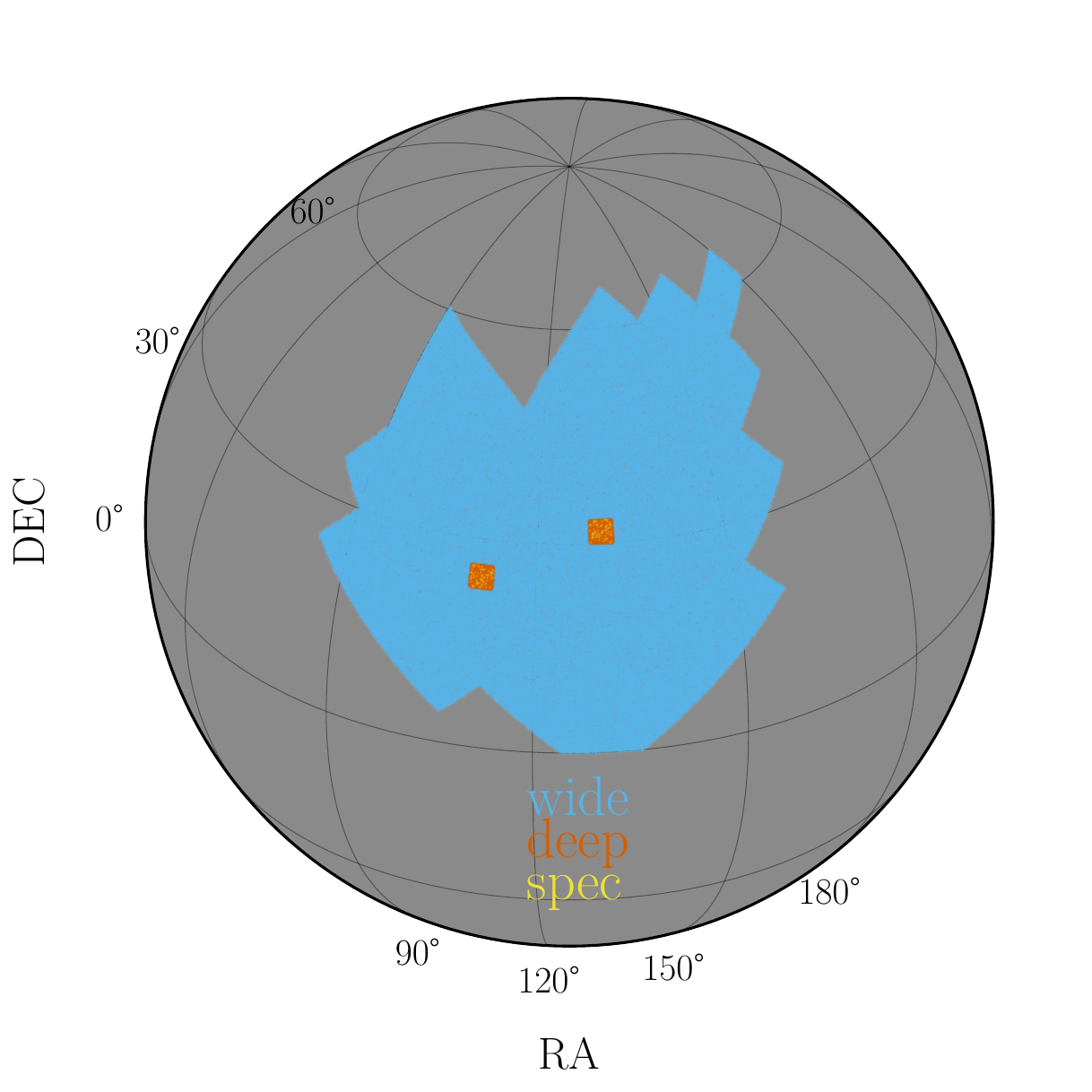}
    	\caption{Sky footprints of the synthetic Roman galaxy catalogs from the \texttt{Cardinal} simulations. The adopted areas of the deep and wide fields (HLIS deep and medium tiers) are 19.24~$\text{deg}^2$ and 2000~$\text{deg}^2$, respectively. The COSMOS-matched spec-$z$ catalog (yellow) is a subset of the deep-field photometric catalog (orange) with known redshifts. The catalogs are downsampled proportionally here for visualization. }
        \label{fig:footprint}
    \end{figure}
\end{center}

The Vera C. Rubin Observatory Legacy Survey of Space and Time (Rubin LSST) is expected to observe the same deep fields in the $u$, $g$, $r$, $i$, $z$, and $y$ optical bands. Cosmological analyses are expected to combine Roman photometry with Rubin forced-photometry measurements \citep{slimfarmer}. We simulate four years of LSST observations using the observing strategy specified by OpSim v5.1.0, matching the expected timeline for Roman medium-tier data. According to the current observing schedule,\footnote{\url{https://roman-docs.stsci.edu/roman-community-defined-surveys/roman-observations-in-the-first-two-years-of-science-operations}} the deep tier is expected to provide the first weak lensing sample before the full medium-tier survey is complete. We therefore adopt the \texttt{Cardinal} deep configuration for the forecasts of an early Roman weak lensing analysis in Section~\ref{sec:cosmo}. Figure~\ref{fig:footprint} shows footprints of the simulated catalogs. 

To emulate spectroscopic selection, we match \texttt{Cardinal} galaxies to the COSMOS spectroscopic compilation of \citet{Khostovan2026}. 
We first select COSMOS galaxies with reliable spectroscopic redshifts ($Q=3$ or $4$) and match them to the \texttt{COSMOS2020} \citep{COSMOS2020} classic catalog to obtain photometry in the $u$, $g$, $r$, $i$, $z$, $y$, $Y$, $J$, and $H$ bands. We retain sources with valid measurements in all bands and redshifts $z<2.3$. For the \texttt{Cardinal} galaxies, we perturb the true fluxes using an empirical noise model derived from \texttt{COSMOS2020}. In each band, we assign a flux uncertainty drawn from one of the 100 \texttt{COSMOS2020} galaxies with the most similar observed fluxes and add a corresponding Gaussian noise realization to the simulated true flux. We then match each spec-$z$ galaxy with a unique \texttt{Cardinal} galaxy within redshift and $H$-band magnitude bins of widths $\Delta z=0.05$ and $\Delta H=0.1$, respectively. We use $H$-band magnitude as the brightness coordinate for matching because $H$ will be the main reference band for Roman weak lensing source selection. 
Within each bin, we perform nearest-neighbor matching using the percentile ranks of eight colors: $u-g$, $g-r$, $r-i$, $i-z$, $z-y$, $y-Y$, $Y-J$, and $J-H$. Using ranks rather than absolute colors reduces sensitivity to differences between the observed and simulated color distributions. 

Finally, we apply $\mathrm{S/N}_{YJH}>18$ cuts to all three catalogs and remove galaxies with $\mathrm{mag}\geq30$ in any deep band to exclude extreme outliers arising from simulation artifacts. These cuts leave 28,583 galaxies in the COSMOS-matched, realistically incomplete spec-$z$ sample, hereafter referred to as the COSMOS-matched spec-$z$. The final deep-field and wide-field samples for photo-$z$ characterization contain 3,446,009 and 66,234,521 galaxies, respectively.

\section{SOMPZ Methodology} 
\label{sec:sompz}

This section reviews the standard DES-SOMPZ formalism and examines its limitations under incomplete spectroscopic coverage. We apply DES-SOMPZ to the simulated Roman catalogs, using the implementation in the Redshift Assessment Infrastructure Layers \citep[RAIL;][]{RAIL} software library.\footnote{\url{https://github.com/LSSTDESC/RAIL}} The Roman-specific redshift characterization and uncertainty propagation workflow is maintained in the HLIS Cosmology Project Infrastructure Team's \texttt{Roman-SOMPZ} repository.\footnote{\url{https://github.com/Roman-HLIS-Cosmology-PIT/Roman-SOMPZ}}

SOMPZ is a redshift characterization framework widely used in photometric galaxy surveys \citep{KIDSsompz, SOMPZ-DES_Y3,2024MNRAS.527.2010G, SOMPZ-DES_Y6, Euclidsompz, 2026PhRvD.114b3520G}. In DES analyses, SOMPZ uses two-stage self-organizing maps (SOMs) to connect three input catalogs: a spectroscopic (spec) catalog 
containing galaxies with high-confidence redshifts, a deep-field (deep) catalog with deep multiband photometry, and a wide-field (wide) catalog with shallower photometry in fewer bands that defines the weak lensing source sample. By linking the redshifts and photometric phenotypes of galaxies across these catalogs, SOMPZ estimates the ensemble redshift distribution of the wide-field sample in each tomographic bin.

SOMPZ divides the wide-field source population into phenotypes and assigns each phenotype to a tomographic bin. It then estimates the $n(z)$ in each bin by reweighting the spec-$z$ distribution based on the abundance of each phenotype. A SOM represents galaxy multiband photometry on a two-dimensional grid, with each cell corresponding to a color phenotype. 
We train separate SOMs using the deep-field and wide-field photometric catalogs and  denote their cells by $c$ and $\hat{c}$, respectively. Each wide-SOM cell $\hat{c}$ is assigned to the tomographic bin $\hat{b}$ containing most of 
its spec-$z$ galaxies. SOMPZ then estimates the bin's redshift distribution via equation 
\begin{equation}
    n(z|\hat{b}) \propto \sum_{\hat{c} \in \hat{b}} \sum_c p(z|c, \hat{b})p(c|\hat{c})p(\hat{c}).
    \label{eq:sompz}
\end{equation}
The first term on the right-hand side, $p(z|c,\hat{b})$, is the conditional distribution of spec-$z$ in deep-SOM cell $c$ and tomographic bin $\hat{b}$. The transfer function $p(c|\hat{c})$ links the deep and wide SOMs and is estimated using the deep catalog, which has both simulated deep- and wide-field photometry available and thus cell assignments of both SOMs. The final term, $p(\hat{c})$, is estimated from the fraction of source population in wide-SOM cell $\hat{c}$. Following \cite{SOM-DES_Y6}, when no redshift-reference galaxy is assigned to a particular bin $\hat{b}$ in a given deep-SOM cell $c$ to evaluate $p(z|c, \hat{b})$, we relax the bin conditioning and replace the term with $p(z|c)$, using all redshift-reference galaxies in cell $c$.

\begin{center}
    \begin{figure*}[!t]
        \centering
    	\includegraphics[width=\linewidth]{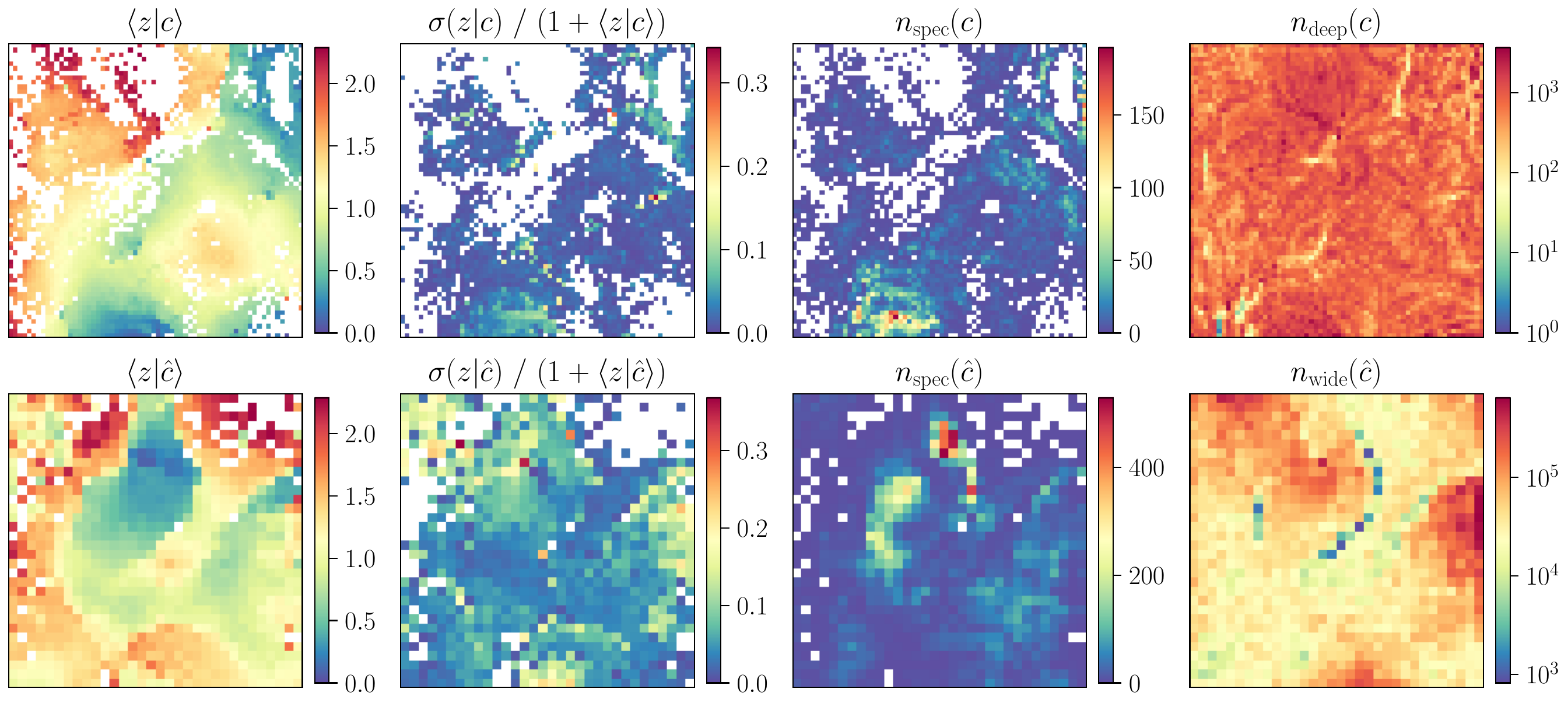}
    	\caption{
        Self-organizing maps (SOMs) trained using the Roman-like deep-field (top row) and wide-field (bottom row) photometry. 
        The deep SOM has 64\texttimes64 cells ($c$) and the wide SOM has 32\texttimes32 cells ($\hat{c}$). The SOM cells are colored by the mean spectroscopic redshift (spec-$z$; leftmost column), the normalized standard deviation of spec-$z$ (second column), the number of spec-$z$ galaxies (third column), and the number of deep-field (top-right) or wide-field (bottom-right) galaxies. For each row, the white cells in the leftmost panel (or similarly in the middle two panels) are what we refer to as ``no-spec cells", i.e., SOM cells with no spec-$z$ galaxies assigned to them. These no-spec deep- and wide-SOM cells contain 26.6\% and 7.2\% of the deep- and wide-field galaxies, respectively. }  
        \label{fig:SOMs}
    \end{figure*}
\end{center}

Figure~\ref{fig:SOMs} shows the deep and wide SOMs trained on the \texttt{Cardinal} catalogs. From left to right, the SOM cells are colored by mean redshift, redshift scatter, and the numbers of spectroscopic and photometric galaxies. The mean redshift varies smoothly across most cells in each map, with a few sharp discontinuities corresponding to regions of color--redshift degeneracy, as seen in previous studies \citep[e.g.,][]{Masters15, SOMPZ-DES_Y3, UMAP-kNN-z, SOM-DES_Y6}. Most cells have small redshift scatter; the scatter increases in some cells near those mean-redshift discontinuities
or the boundaries of the maps. 

However, incomplete spec-$z$ coverage of the SOMs (Fig.~\ref{fig:SOMs}, leftmost column) introduces two limitations in our application of DES-SOMPZ to the Roman-like catalogs: 

\begin{enumerate}
\item During tomographic binning, DES-SOMPZ excludes the $7.2\%$ of source galaxies in wide-SOM cells without spec-$z$ galaxies (hereafter ``no-spec cells''); 
\item During redshift calibration, the spec-$z$ reference sample underrepresents faint, high-redshift galaxies, biasing the $n(z)$ estimates in the highest-redshift tomographic bins.
\end{enumerate}

\begin{center}
    \begin{figure}[!t]
        \centering
    	\includegraphics[width=\linewidth]{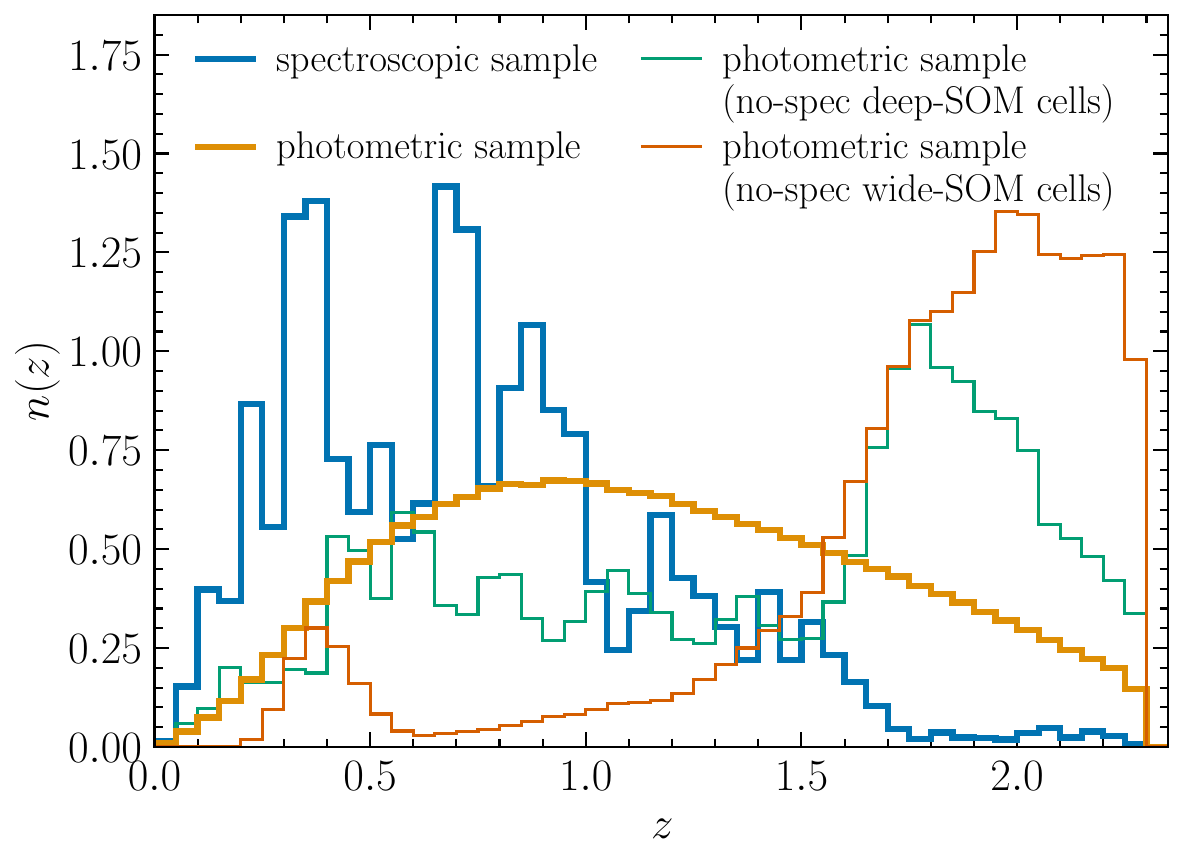}
    	\caption{Redshift distributions of the COSMOS-matched spec-$z$ (blue) and full \texttt{Cardinal} wide-field photometric sample (yellow). Green and red lines represent redshift distributions of the deep-field galaxies in no-spec deep-SOM cells and wide-field galaxies in no-spec wide-SOM cells, respectively. 
        Relative to the full photometric sample, spectroscopic selection skews the spec-$z$ reference sample toward low-redshift galaxies, 
        whereas galaxies in no-spec SOM cells preferentially lie at higher redshifts. 
        } 
        \label{fig:z_dists}
    \end{figure}
\end{center}

The second limitation arises from the more incomplete spec-$z$ coverage of the deep SOM, where $26.6\%$ of deep-field galaxies are left in no-spec cells. These no-spec cells preferentially contain faint, high-redshift photometric galaxies, yet no spec-$z$ available to provide estimates for their $p(z|c,\hat{b})$ distributions. Through the transfer function $p(c|\hat{c})$, this missing high-redshift representation also affects the estimated redshift distributions of the wide-field sample, biasing $n(z)$ 
particularly in the highest-redshift tomographic bins. Figure~\ref{fig:SOMs} top-left panel shows the positions of these no-spec cells. Figure~\ref{fig:z_dists} compares the spec-$z$ distribution with the redshift distributions of the full photometric sample and the subsets in no-spec cells of either SOM. 

These two limitations motivate modifications to the tomographic binning and redshift estimation of DES-SOMPZ. We next review approaches to mitigate spectroscopic data incompleteness and propose the Roman-SOMPZ extensions that address incomplete spec-$z$ at each stage.

\section{Roman-SOMPZ Advancement}
\label{sec:roman_sompz}

Mitigating biases in photo-$z$ characterization caused by incomplete or non-representative spectroscopic reference samples remains a major challenge \citep[see, e.g.,][for a review]{photoz-review}. Current weak lensing surveys have generally pursued two approaches. The first is to improve the completeness and representativeness of the redshift-reference sample by incorporating deep, multiband photometric data, including medium and narrow bands. For example, DES Y6 analyses \citep{SOMPZ-DES_Y6} use additional redshift-reference samples drawn from \texttt{PAUS} \citep{PAUS} and \texttt{COSMOS2020} \citep{COSMOS2020}. However, photo-$z$ catalogs from external surveys can themselves be biased, especially for faint or high-redshift galaxy populations. Those biases can propagate into cosmological constraints, while increased photo-$z$ scatter at faint magnitudes may further limit the characterization. 

The second approach obtains more complete redshift-reference samples through dedicated spectroscopic follow-up programs. Examples include the Complete Calibration of the Color-Redshift Relation survey \citep[C3R2;][]{C3R21,C3R22}, DESI C3R2 \citep{DC3R2}, and the proposed DESI-Deep Survey \citep{DESI-Deep_pilot} and Subaru-PFS/Roman (SuPR) Deep Survey. These programs can improve spectroscopic coverage, but assembling sufficiently complete samples requires considerable observing time, typically at a scale of several years.

We adopt the first approach with modification, distinguishing two tasks: 
\textbf{tomographic binning} assigns source galaxies to tomographic bins, and \textbf{redshift calibration} estimates the ensemble $n(z)$ distribution of the galaxies within each bin. These tasks place different requirements on their respective redshift-reference sample. Tomographic binning primarily requires reliable \textit{relative} redshift information to sort galaxy phenotypes. Redshift calibration instead requires unbiased information about their \textit{absolute} redshifts. 

Based on this distinction, we design two Roman-SOMPZ extensions: improved tomographic bin construction and augmentation of the redshift-reference sample. 
Sections~\ref{sec:tomobin} and \ref{sec:umap} describe each methodological component, respectively. We test the proposed Roman-SOMPZ framework using the simulated photometric catalogs and COSMOS-matched spec-$z$ from \texttt{Cardinal}.

\subsection{Improved tomographic bin construction}
\label{sec:tomobin}
We improve DES-SOMPZ's tomographic bin construction with two objectives: to assign no-spec wide-SOM cells to bins, retaining the source galaxies that the baseline method excludes; and to refine the bin edges to reduce overlaps between the redshift distributions of different tomographic bins, providing better-separated source samples for subsequent weak lensing cosmological analyses. 

Assigning no-spec wide-SOM cells requires redshift estimates for their associated galaxy phenotypes. At this stage, we use these estimates only to rank and bin the cells, not to reconstruct their redshift distributions. The photo-$z$ accuracy requirements 
are therefore less restrictive than would be for redshift estimation. We use photo-$z$ point estimates from \texttt{FlexZBoost} \citep{FlexZBoost, FlexZBoost1}, a gradient-boosted conditional density estimator \citep{XGBoost} implemented in RAIL. With a representative training sample, \texttt{FlexZBoost} ranked among the best-performing methods in the LSST-DESC photo-$z$ data challenge \citep{DC1}. 

We train \texttt{FlexZBoost} on the COSMOS-matched spec-$z$ sample and apply it to the deep catalog. Despite incomplete spec-$z$, low-noise deep-field photometry enables precise individual-galaxy photo-$z$ estimation. Directly training and applying the estimator with the noisier wide-field photometry does not achieve comparable performance. 

\begin{center}
    \begin{figure}[!t]
        \centering
    	\includegraphics[width=\linewidth]{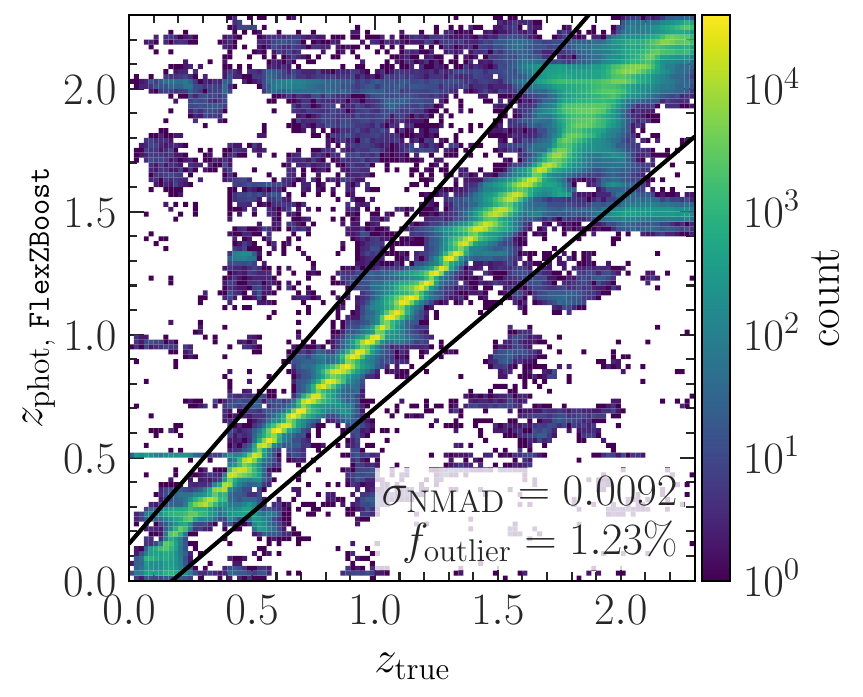}
    	\caption{\texttt{FlexZBoost} estimated vs. true redshifts of \texttt{Cardinal} deep-field galaxies. Point estimates are the medians of the predicted individual galaxy redshift distributions, and colors indicate galaxy counts on a logarithmic scale. The concentration of galaxies along the diagonal indicates precise redshift estimates for most of the sample. }
        \label{fig:FlexZBoost}
    \end{figure}
\end{center}

Figure~\ref{fig:FlexZBoost} compares the \texttt{FlexZBoost} estimates for the deep-field galaxies with their true redshifts. Each point estimate is defined as the median of the galaxy's predicted $p(z)$ distribution. 
Most galaxies lie close to the diagonal, indicating low photo-$z$ scatter that supports the use of these estimates for tomographic bin construction. 
The outliers have little impact on tomographic binning because each SOM cell's assignment follows the mode of its cell-level redshift distribution. 

To integrate our tomographic bin construction method into the SOMPZ pipeline in a consistent way, we first combine the spec-$z$ sample with \texttt{FlexZBoost} estimates for the rest of the deep-field galaxies. 
This hybrid reference sample is then used to determine the bin edges and wide-SOM cell assignments. 
We emphasize that the \texttt{FlexZBoost} point estimates are used only for binning here and they do not enter the subsequent $n(z)$ estimation using Eq.~\ref{eq:sompz}. This is because to obtain an unbiased $n(z)$ needed for weak lensing sources, more emphasis would be placed on the accuracy than on the precision of the redshift-reference sample. As we show later, the accuracy of \texttt{FlexZBoost} is not sufficient for unbiased $n(z)$ estimation. 

The top row of Fig.~\ref{fig:Nz} compares DES-SOMPZ characterization of the \texttt{Cardinal} wide catalog using $\sim20$k representative spec-$z$'s across the two $9.62~\rm{deg}^2$ deep fields (top-left) with characterization using the COSMOS-matched spec-$z$ (top-right). 
Relative to the ideal case, incomplete spec-$z$ results in greater overlaps among redshift distributions of the low-redshift bins. This is because DES-SOMPZ defines bin edges from the quantiles of the spec-$z$ distribution. The COSMOS-matched spec-$z$ distribution is right skewed (Fig.~\ref{fig:z_dists} blue line), and its quantiles place the DES-SOMPZ bin edges close together at low redshift. 
We instead define the bin edges using the combined spec-$z$ and deep photo-$z$ sample, which better represents the wide-field population. We further replace the quantiles with Jenks natural breaks \citep{Jenks}. This method 
minimizes the total within-bin redshift variance and thereby maximizes the deviations between tomographic bins. 

Note that we adopt nine tomographic bins throughout the analyses. The Roman fiducial configuration uses ten bins \citep{Eifler21}, but we omit the highest-redshift bin because \texttt{Cardinal} contains no galaxies above $z\simeq 2.3$. This cutoff may have two implications. First, the omitted highest-redshift bin is expected to have little spec-$z$ coverage, so our analysis may underestimate the challenge for characterizing the full ten-bin Roman sample. Second, the simulation cutoff at $z\simeq2.3$ creates an artificial sharp upper edge in the true-$z$ distribution of the retained highest-redshift bin (bin~9), which is difficult for machine learning-based photo-$z$ estimators to reproduce. Discrepancies near it therefore reflect truncation of the simulated population rather than the behavior expected for a source sample extending to higher redshift. This artifact will not be present in real observations, but the additional high-redshift population will introduce photo-$z$ characterization challenges that we do not test here. 

Table~\ref{tab:binning} summarizes the tomographic bin construction methods in DES-SOMPZ and our proposed Roman-SOMPZ. Since the baseline pipeline already assigns deep-field galaxies to wide-SOM cells, our extension adds only the computational cost of producing the \texttt{FlexZBoost} photo-$z$ estimates and computing the Jenks natural breaks.

\begin{center}
    \begin{figure*}[!t]
        \centering
    	\includegraphics[width=\linewidth]{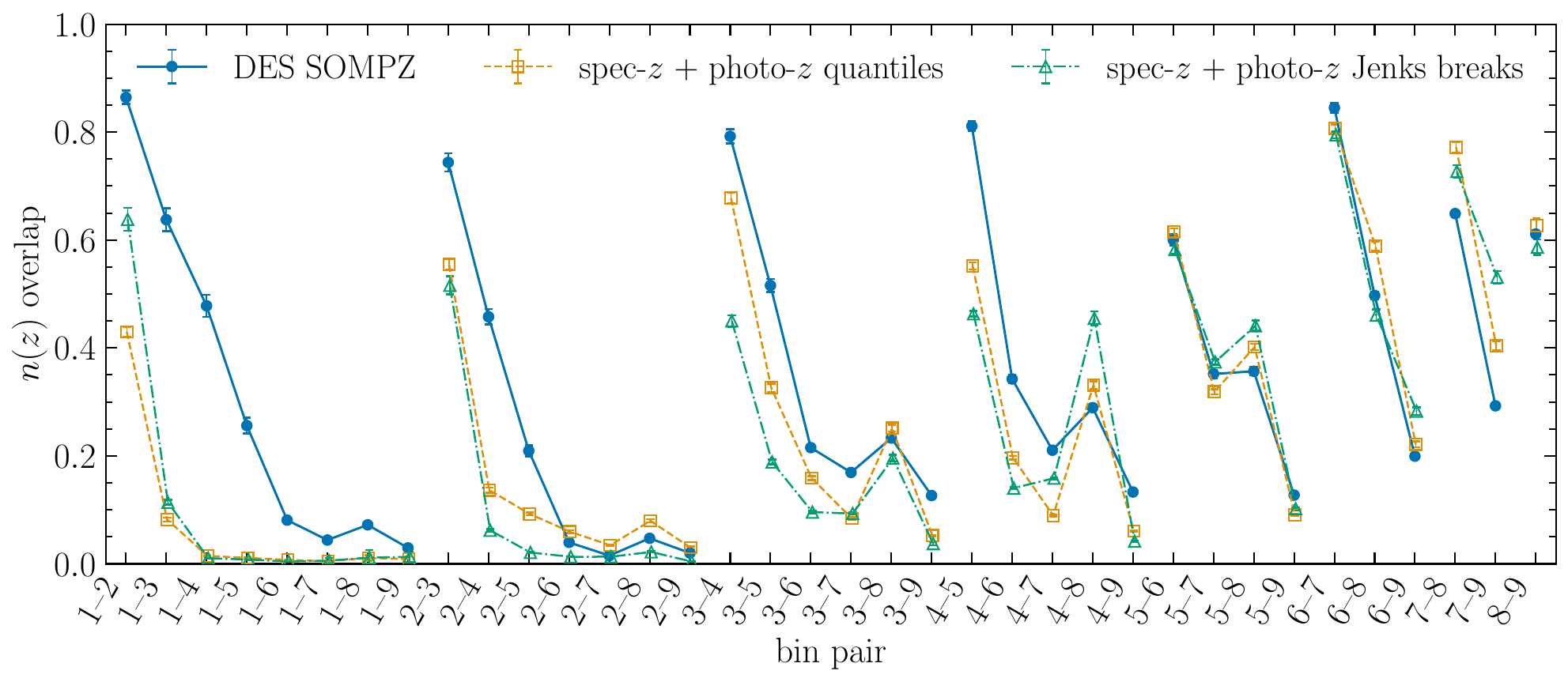}
    	\caption{$n(z)$ overlaps between all pairs of tomographic bins for three binning methods, quantified by cosine similarity (Eq.~\ref{eq:bin_overlaps}). 
        Blue circles correspond to the baseline DES-SOMPZ binning based on spec-$z$ quantiles. Yellow squares and green triangles use the combined spec-$z$ and \texttt{FlexZBoost} photo-$z$ sample, with bin edges defined by quantiles and Jenks natural breaks, respectively. Incorporating \texttt{FlexZBoost} photo-$z$'s reduces overlap for most bin pairs relative to DES-SOMPZ, with Jenks natural breaks generally providing a further reduction. Error bars are estimated from $n(z)$ realizations. }
        \label{fig:bin_overlaps}
    \end{figure*}
\end{center}

\begin{table}
  \centering
   \nolinenumbers  
    \caption{Summary of the tomographic bin construction methods used in DES-SOMPZ and Roman-SOMPZ.}
  \renewcommand{\arraystretch}{1.2}
  \begin{tabularx}{\columnwidth}{l >{\raggedright\arraybackslash}X >{\raggedright\arraybackslash}X}
    \toprule
    Method & Bin assignment (cell-level) & Bin edges \\
    \midrule
    DES-SOMPZ & based on the majority of spec-$z$; no-spec cells excluded & spec-$z$ quantiles \\
    \addlinespace
    Roman-SOMPZ & based on the majority of combined spec-$z$ \& photo-$z$; applies to all SOM cells & spec-$z$ \& photo-$z$ Jenks natural breaks \\
    \bottomrule
  \end{tabularx}
  \label{tab:binning}
\end{table}
To compare the binning strategies, we measure the overlap between the redshift distributions of each pair of tomographic bins \citep{SOM-DES_Y6}. We represent the $n(z)$ distributions as normalized galaxy-count vectors on a common redshift grid and define their overlap as the cosine similarity: 
\begin{equation}
O_{i, j}
=
\frac{\mathbf{n}_i \cdot \mathbf{n}_j}
{\lVert\mathbf{n}_i\rVert \lVert\mathbf{n}_j\rVert}, 
    \label{eq:bin_overlaps}
\end{equation}
where $i$ and $j$ denote the bin numbers. Figure~\ref{fig:bin_overlaps} shows that DES-SOMPZ produces the largest overlaps for most bin pairs. Incorporating \texttt{FlexZBoost} photo-$z$'s alone reduces these overlaps, and Jenks natural breaks generally reduce them further. 

The bottom-left panel of Fig.~\ref{fig:Nz}  shows the $n(z)$ estimates of the \texttt{Cardinal} wide catalog after we apply the improved tomographic binning. Although this step recovers galaxies in no-spec wide-SOM cells and yields better-separated bins, the highest-redshift bins remain biased with discrepancies increased relative to DES-SOMPZ. 
Galaxies in no-spec wide-SOM cells typically lie at higher redshifts than the spec-$z$ sample (Fig.~\ref{fig:z_dists}, red vs. blue lines). Similarly, they are more likely to correspond to no-spec deep-SOM cells as well. For these no-spec deep-SOM cells, $p(z|c)$ is unavailable in Eq.~\ref{eq:sompz}. 
The transfer function $p(c|\hat{c})$ and summation over deep-SOM cells propagates the underrepresentation of high-redshift phenotypes to the estimated $n(z)$. This biases the
distributions toward lower redshift, especially in the highest-$z$ bins where no-spec cells are the most prevalent. We therefore must also supply redshift information for no-spec deep-SOM cells, which is our next major design step in Roman-SOMPZ. 

\subsection{Training sample augmentation}
\label{sec:umap}
To fill in the no-spec deep-SOM cells, a natural thought would be to interpolate redshifts directly across the deep SOM. However, projecting multiband color space onto a square grid distorts distances between galaxies. The sharp redshift discontinuities associated with color--redshift degeneracy (Fig.~\ref{fig:SOMs}, top-left; see discussion in Section~\ref{sec:sompz}) further complicate interpolation. 
The Uniform Manifold Approximation and Projection \citep[UMAP;][]{UMAP} algorithm offers a continuous, compressed color space designed to preserve local and global structure of the data. \cite{UMAP-kNN-z} illustrated that a spec-$z$ sample populates a thin manifold in a three-dimensional UMAP embedding, along which redshift varies monotonically. This structure allows for accurate redshift regression even in sparsely sampled regions, including at high redshift. We therefore use regression in UMAP space as a more principled approach to augmenting the Roman-SOMPZ redshift-reference sample. 

We first construct a three-dimensional UMAP embedding of the deep catalog from seven colors ($u-g$, $g-r$, $r-i$, $i-z$, $z-y$, $y-J$, and $J-H$) that is consistent with the input features in \cite{UMAP-kNN-z}. We then train a normalizing flow \citep{flow-MAF, flow-transformer} on the simulated spec catalog to model the joint distribution of redshift and UMAP coordinates. Normalizing flows are flexible generative models capable of approximating complex, high-dimensional distributions, which motivates their use for estimating this joint density with the training sample being highly non-uniform. 

For each deep-field galaxy in a no-spec cell, we evaluate the learned joint density at its UMAP coordinates and normalize the resulting redshift slice to obtain a $p(z)$ posterior. We name this estimator UMAP-flow-$z$ and adopt the posterior medians as its photo-$z$ point estimates, which yields the smallest photo-$z$ bias among the posterior mean, median, and mode. 
Appendix~\ref{app:umap-flow-z} describes the method and implementation in detail. 

\begin{center}
    \begin{figure}[!t]
        \centering
    	\includegraphics[width=\linewidth]{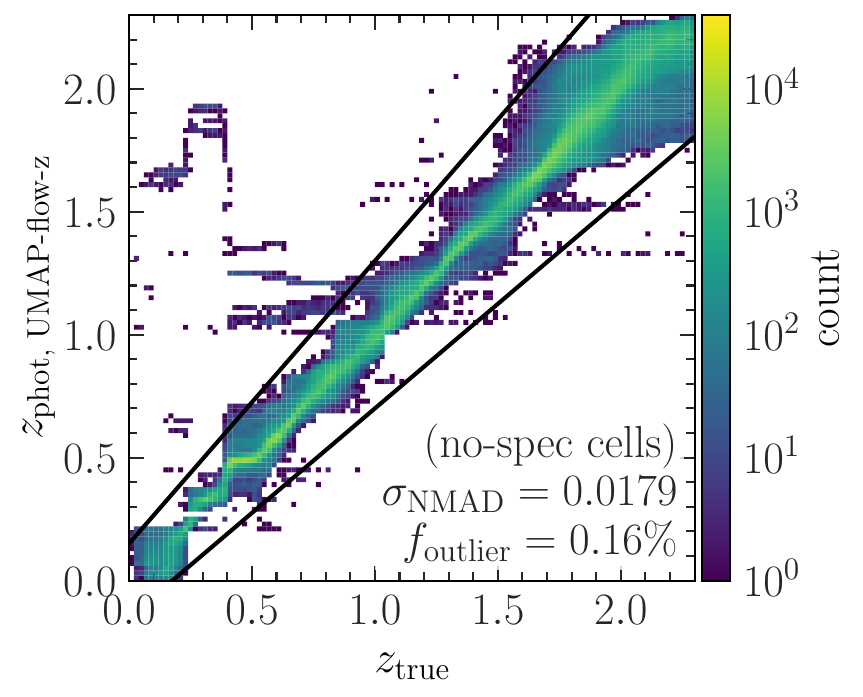}
    	\caption{UMAP-flow-$z$ estimated vs. true redshifts of \texttt{Cardinal} deep-field galaxies in no-spec deep-SOM cells. Point estimates are the medians of the predicted individual galaxy redshift distributions, and colors indicate galaxy counts on a logarithmic scale. The estimates follow the true redshifts with a low outlier fraction. 
        This motivates their use to augment the redshift-calibration sample, where outliers can bias the estimated ensemble $n(z)$ distributions. 
        }
        \label{fig:UMAP-flow-z}
    \end{figure}
\end{center}

Figure~\ref{fig:UMAP-flow-z} compares the UMAP-flow-$z$ estimates with true redshifts for the deep-field galaxies in no-spec cells. The estimates have very few extreme outliers, although more scattered about the diagonal than the \texttt{FlexZBoost} estimates shown in Fig.~\ref{fig:FlexZBoost}. 
An additional evaluation of \texttt{FlexZBoost} within the no-spec cells confirms that UMAP-flow-$z$ has a larger normalized median absolute deviation ($\sigma_\text{NMAD}$) scatter but lower outlier fraction. 
This trade-off motivates our use of UMAP-flow-$z$ for deep-SOM augmentation, as the \texttt{FlexZBoost} outlier populations may become dominant in the calibrated $n(z)$s of the highest-redshift bins. 

\begin{center}
    \begin{figure*}[!t]
        \centering
    	\includegraphics[width=\linewidth]{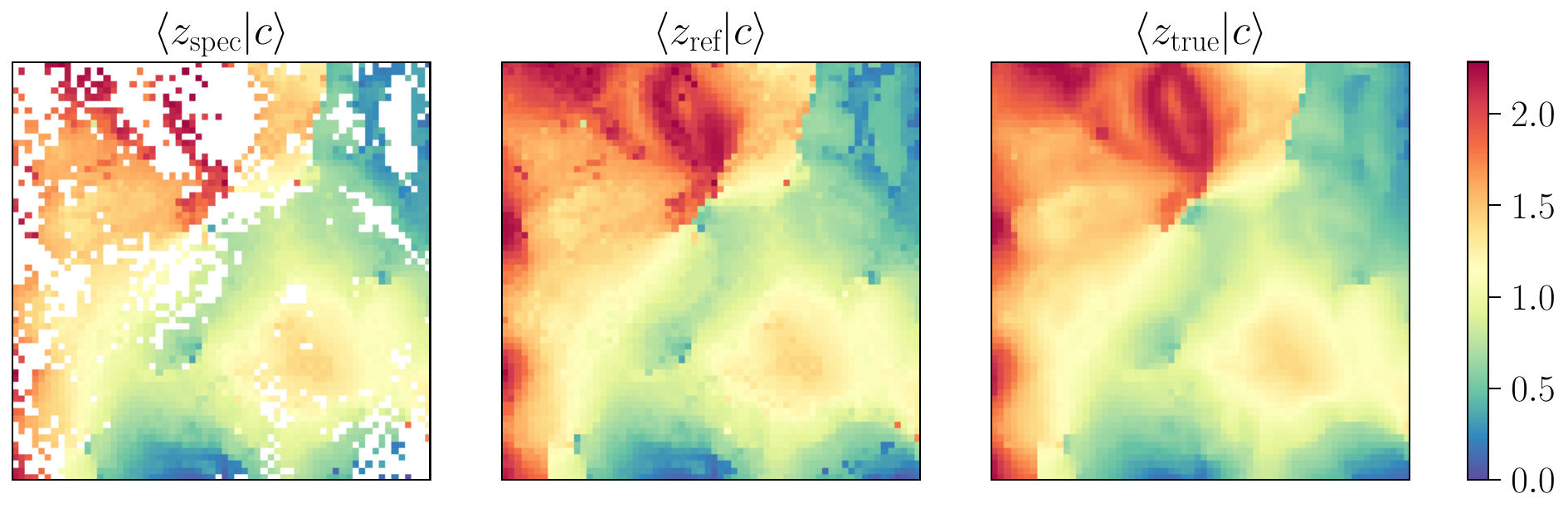} 
    	\caption{Deep SOM with cells colored by mean redshift of the calibration sample before (left) and after (middle) augmentation, compared with the true redshifts (right). 
        \textit{Left:} COSMOS-matched spec-$z$; same as the top-left panel of Fig.~\ref{fig:SOMs}. 
        \textit{Middle:} Reference redshifts after augmentation with UMAP-flow-$z$ estimates for galaxies in no-spec deep-SOM cells (white cells in the left panel). 
        \textit{Right:} Mean true redshift of the \texttt{Cardinal} deep galaxies in each cell. 
        The augmentation supplies redshift information for no-spec cells while preserving mean-redshift trends across the SOM and capturing the structure of the true mean-redshift map. }
        \label{fig:SOM-augmented}
    \end{figure*}
\end{center}

We augment the spec-$z$ reference sample with these UMAP-flow-$z$ estimates and follow Eq.~\ref{eq:sompz} to estimate the wide-field $n(z)$ distributions. The middle panel of Fig.~\ref{fig:SOM-augmented} shows the deep SOM colored by mean redshift of the augmented reference sample in each cell. 
Comparing with the left panel demonstrates how augmentation fills the spectroscopically empty cells while preserving the mean-redshift trends across the map. The right panel shows the true cell-averaged redshifts
for a visual comparison. The UMAP manifolds populated by the corresponding samples are presented in Appendix~\ref{app:umap-flow-z-prediction}. 

\subsection{Performance and interpretation}

\begin{center}
    \begin{figure*}[!t]
        \centering
    	\includegraphics[width=\linewidth]{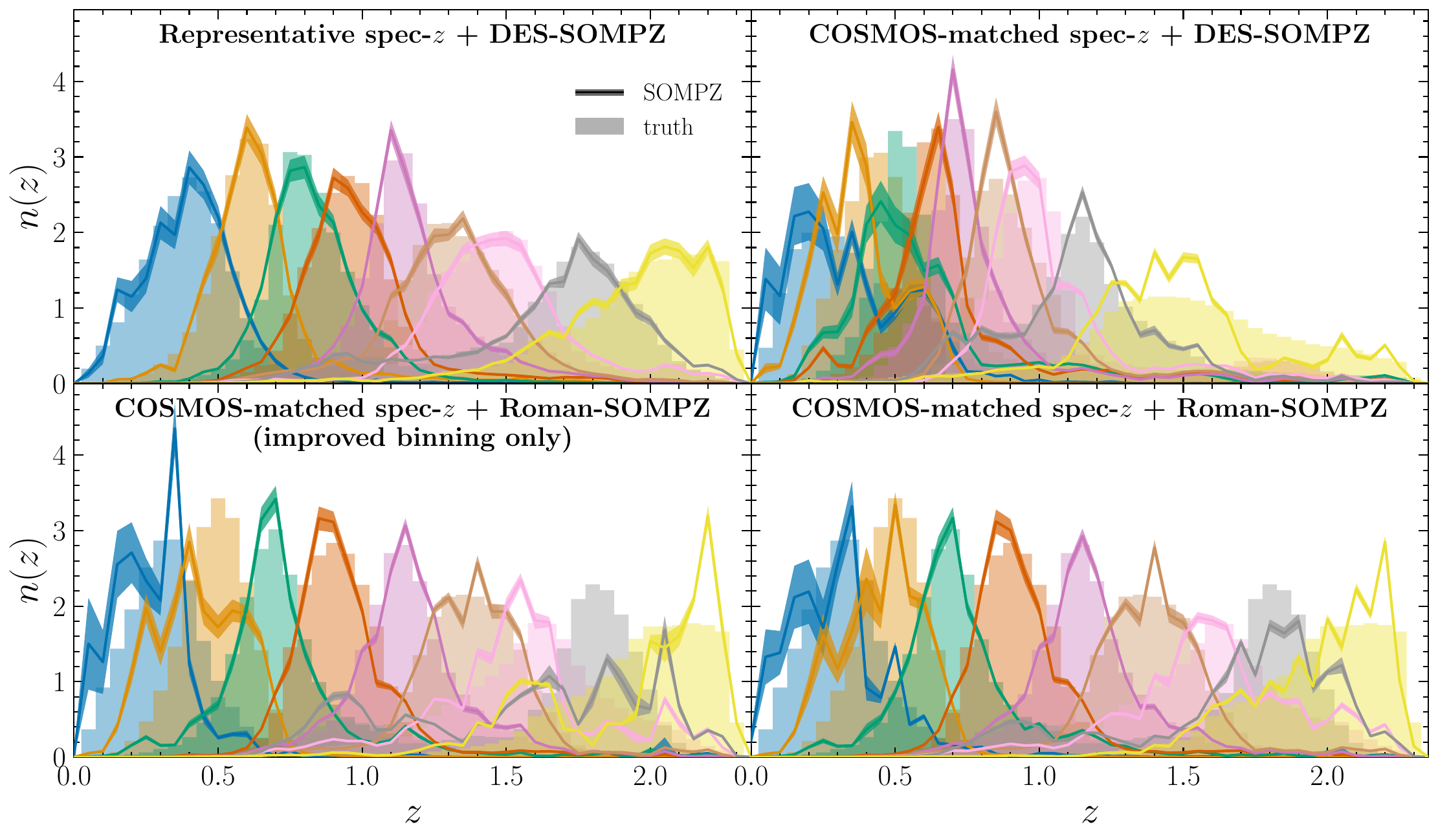}
        \caption{
SOMPZ redshift characterization of the \texttt{Cardinal} wide catalog. Colors distinguish the nine tomographic bins; solid lines show the SOMPZ estimates, and semi-transparent histograms show the true redshift distributions of the corresponding galaxies assigned to each bin. Uncertainty bands around the lines indicate the standard deviation across 100 $n(z)$ realizations including sample variance and shot noise for the adopted $19.24\,\mathrm{deg}^2$ deep-field area.
\textit{Top-left vs. top-right:} Replacing the representative spectroscopic reference sample with the COSMOS-matched spec-$z$ sample for DES-SOMPZ produces more overlapping tomographic bins and larger discrepancies between the estimated and true $n(z)$ distributions, particularly at high redshift.
\textit{Top-right vs. bottom-left:} Using the same COSMOS-matched spec-$z$ sample, Roman-SOMPZ's improved tomographic bin construction method (Section~\ref{sec:tomobin}) recovers galaxies in previously unassigned wide-SOM cells and reduces overlaps between bins. However, missing redshift information in no-spec deep-SOM cells still biases the inferred $n(z)$ distributions toward lower redshift in the highest-$z$ bins (bin 7-9).
\textit{Bottom-left vs. bottom-right:} 
The full Roman-SOMPZ framework retains the improved tomographic binning and augments the redshift-calibration sample with UMAP-flow-$z$ estimates for galaxies in no-spec deep-SOM cells (Section~\ref{sec:umap}). This brings the estimated $n(z)$ distributions into closer agreement with the truth, particularly in the highest-redshift bins, although residual biases remain.
        } 
        \label{fig:Nz}
    \end{figure*}
\end{center}

The bottom-right panel of Fig.~\ref{fig:Nz} shows \texttt{Cardinal} wide-field $n(z)$ distributions characterized by the full Roman-SOMPZ framework, applying all the algorithmic improvements in Sections~\ref{sec:tomobin}~\&~\ref{sec:umap}. Compared with the result from only improved binning (Fig.~\ref{fig:Nz}, bottom-left), UMAP-flow-$z$ training-sample augmentation reduces the bias toward lower-$z$ in the highest-redshift tomographic bins. 

Relative to baseline DES-SOMPZ, Roman-SOMPZ reduces the absolute mean-redshift biases in eight of nine tomographic bins (Table~\ref{tab:mean_redshift}). 
The shapes of the estimated $n(z)$ distributions also more closely resemble those obtained in the idealized case with representative spec-$z$ across the deep fields (Fig.~\ref{fig:Nz}, bottom-right vs. top-left). We have thus far implemented the full framework of Roman-SOMPZ and demonstrated its effectiveness in mitigating spec-$z$ incompleteness through fully exploiting the spectroscopic and deep-field data using multiple, specialized individual-galaxy photo-$z$ estimators.

The two photo-$z$ estimators contribute in different ways. For tomographic binning, \texttt{FlexZBoost} provides precise redshift information to rank phenotypes and assign no-spec wide-SOM cells. Together with Jenks natural breaks, these estimates also produce better-separated tomographic bins (Fig.~\ref{fig:bin_overlaps}). Redshift estimation is more sensitive to outliers, which can bias the ensemble distributions. 
We use UMAP-flow-$z$ to supply reference redshifts for deep-SOM cells without spec-$z$ coverage. 
Its estimates show few outliers, despite having larger scatter than the \texttt{FlexZBoost} estimates. 

To test the choice of augmentation estimator, we repeat the Roman-SOMPZ characterization with the same tomographic binning but use \texttt{FlexZBoost} instead of UMAP-flow-$z$ to estimate reference redshifts for no-spec deep-SOM cells. 
Table~\ref{tab:mean_redshift} shows that the \texttt{FlexZBoost}-augmented configuration performs well in some lower-redshift bins, but produces larger biases in all three highest-redshift bins (bins~7--9) than with UMAP-flow-$z$ augmentation due to the catastrophic outliers. These results support matching each estimator to the task for which its error properties are best suited, rather than choosing one estimator from a global performance metric. 

\begin{table*}
\centering
\footnotesize
\setlength{\tabcolsep}{2.5pt}
\caption{Mean-redshift biases $\Delta\langle z \rangle/(1+\langle z\rangle)$ for each tomographic bin, with uncertainties estimated from $n(z)$ realizations. We compare the per bin mean-redshift biases under the four SOMPZ configurations in Fig.~\ref{fig:Nz}, with an additional \texttt{FlexZBoost}-augmented Roman-SOMPZ configuration where \texttt{FlexZBoost} point estimates provide the additional redshift references for tomographic binning as well as for $n(z)$ estimation. 
The UMAP-flow-$z$-augmented Roman-SOMPZ configuration proposed in Section~\ref{sec:roman_sompz} reduces the absolute mean-redshift biases relative to DES-SOMPZ in eight of nine tomographic bins and relative to the \texttt{FlexZBoost}-augmented Roman-SOMPZ in bins 7-9. Bins meeting the $|\Delta\langle z\rangle|/(1+\langle z\rangle)<0.002$ photo-$z$ calibration requirement specified in Roman's Science Requirements Document \citep[][see HLIS 2.1.10]{RomanSRD} are boldfaced. 
}%
\begin{tabular}{l ccccccccc}
\toprule
$\Delta\langle z \rangle/(1+\langle z\rangle)$ [$\times10^{-3}$] & Bin 1 & Bin 2 & Bin 3 & Bin 4 & Bin 5 & Bin 6 & Bin 7 & Bin 8 & Bin 9 \\
\midrule
DES-SOMPZ
  & \ve{-21.4}{4.7} & \ve{-45.5}{2.8} & \ve{-28.0}{3.4} & \ve{-25.0}{2.6} & \ve{-22.0}{2.0} & \ve{-31.4}{1.7} & \ve{-16.9}{1.6} & \ve{-2.2}{1.6} & \ve{-11.6}{0.9} \\
\makecell[l]{Roman-SOMPZ -- binning only}
  & \ve{-39.6}{15.8} & \ve{-37.4}{2.6} & \ve{-8.3}{2.4} & \ve{-5.1}{1.3} & \ve{\mathbf{1.9}}{\mathbf{1.0}} & \ve{\mathbf{-1.6}}{\mathbf{0.8}} & \ve{-17.5}{1.2} & \ve{-32.1}{2.3} & \ve{-9.8}{1.3} \\
\makecell[l]{Roman-SOMPZ -- \texttt{FlexZBoost}} augmentation 
  & \ve{8.8}{8.6} & \ve{-28.3}{1.6} & \ve{\mathbf{-1.9}}{\mathbf{1.7}} & \ve{\mathbf{-0.9}}{\mathbf{1.0}} & \ve{\mathbf{1.7}}{\mathbf{0.9}} & \ve{\mathbf{0.2}}{\mathbf{1.0}} & \ve{8.7}{1.4} & \ve{16.0}{2.0} & \ve{3.5}{1.3} \\
Roman-SOMPZ -- UMAP-flow-$z$ augmentation 
  & \ve{-15.5}{7.4} & \ve{-26.4}{1.7} & \ve{-4.1}{1.6} & \ve{\mathbf{-1.4}}{\mathbf{1.0}} & \ve{\mathbf{1.0}}{\mathbf{0.8}} & \ve{\mathbf{0.1}}{\mathbf{1.0}} & \ve{7.3}{1.3} & \ve{11.6}{2.0} & \ve{2.1}{1.3} \\
\hline
\makecell[l]{DES-SOMPZ, with representative spec-$z$}
  & \ve{4.8}{1.5} & \ve{2.3}{1.8} & \ve{\mathbf{-1.4}}{\mathbf{1.2}} & \ve{\mathbf{-0.8}}{\mathbf{1.3}} & \ve{-2.4}{1.1} & \ve{-2.2}{1.3} & \ve{\mathbf{-1.8}}{\mathbf{1.3}} & \ve{4.1}{1.9} & \ve{\mathbf{-0.1}}{\mathbf{1.1}} \\
\bottomrule
\end{tabular}
\label{tab:mean_redshift}
\end{table*}

Roman-SOMPZ does not apply a direct photo-$z$ correction to the wide catalog. As in standard SOMPZ, the deep-field sample remains the intermediary between the redshift-reference and the wide-field source samples \citep{SOMPZ,SOMPZ-DES_Y3}. Its higher-quality photometry resolves galaxy populations that remain degenerate in the noisier wide-field measurements, and the transfer function maps the redshift information onto the wide phenotypes. Accordingly, we train the additional photo-$z$ models using the deep-field photometry and apply them to the deep catalog. 
\texttt{FlexZBoost} provides redshift information for bin construction, while UMAP-flow-$z$ supplies missing information of $p(z|c, \hat{b})$. This retains the core SOMPZ architecture while extending its redshift characterization to regions of the color space without spectroscopic coverage.

\subsubsection{Residual bias compared with sample variance and shot noise uncertainty}

The solid lines in Fig.~\ref{fig:Nz} and the mean-redshift offsets in Table~\ref{tab:mean_redshift} are estimates from the unperturbed catalogs and can be used to diagnose algorithmic biases. To assess the biases relative to uncertainty, we generate 100 $n(z)$ realizations that jointly propagate sample variance (SV) and finite-sample shot noise (SN). The uncertainties in Table~\ref{tab:mean_redshift} are the standard deviations of $\Delta\langle z\rangle/(1+\langle z\rangle)$ across the realizations. The realization scheme follows the phenotype-based three-step Dirichlet framework of \citet{SV}, 
which perturbs the abundances of broad superphenotypes (groups of phenotypes with similar redshifts), their redshift fractions, and their constituent deep phenotypes. Redshift-dependent sample variance inflates these Dirichlet fluctuations relative to pure shot noise. The two deep fields are approximated as a single circular top-hat with area $19.24~\mathrm{deg}^{2}$. %
For DES-SOMPZ, the discrepancy of the estimated mean redshift from truth exceeds the $n(z)$ realization scatter in most tomographic bins. 
In bin~2, for example, the normalized mean-redshift bias is $-45.5\times10^{-3}$, whereas the standard deviation across realizations is $2.8\times10^{-3}$. 
For Roman-SOMPZ, the SV+SN uncertainty bands nevertheless do not absorb all residual discrepancies. Bin~6  has a mean-redshift offset smaller than one standard deviation of the $n(z)$ realizations; bins~4, 5, and~9 lie within two standard deviations, while the remaining bins retain larger offsets, especially bins~2, 7, and~8. 

These uncertainties do not include the contribution from the augmentation procedure. We hold the UMAP-flow-$z$ point estimates fixed when generating the $n(z)$ realizations. 
Each galaxy in the augmented reference sample enters as a unit-weight object, and the realization pipeline draws probability realizations conditioned on their joint counts by redshift bin and deep-SOM cell. It does not model the uncertainties in the photo-$z$ estimates themselves. 
The quoted errors therefore do not include uncertainties from retraining the UMAP embedding or normalizing flow and from sampling the predicted individual-galaxy redshift posteriors. They also exclude photometric-calibration uncertainty and imperfections in the simulated galaxy population. Table~\ref{tab:mean_redshift} should thus not be interpreted as a complete Roman redshift-calibration error budget. Quantifying the uncertainty inherent to machine learning-based augmentation is a priority for future work.

\section{Impact of the Roman-SOMPZ Framework on Cosmological Inference}
\label{sec:cosmo}

    \begin{figure}[!t]
        \centering
    	\includegraphics[width=\linewidth]{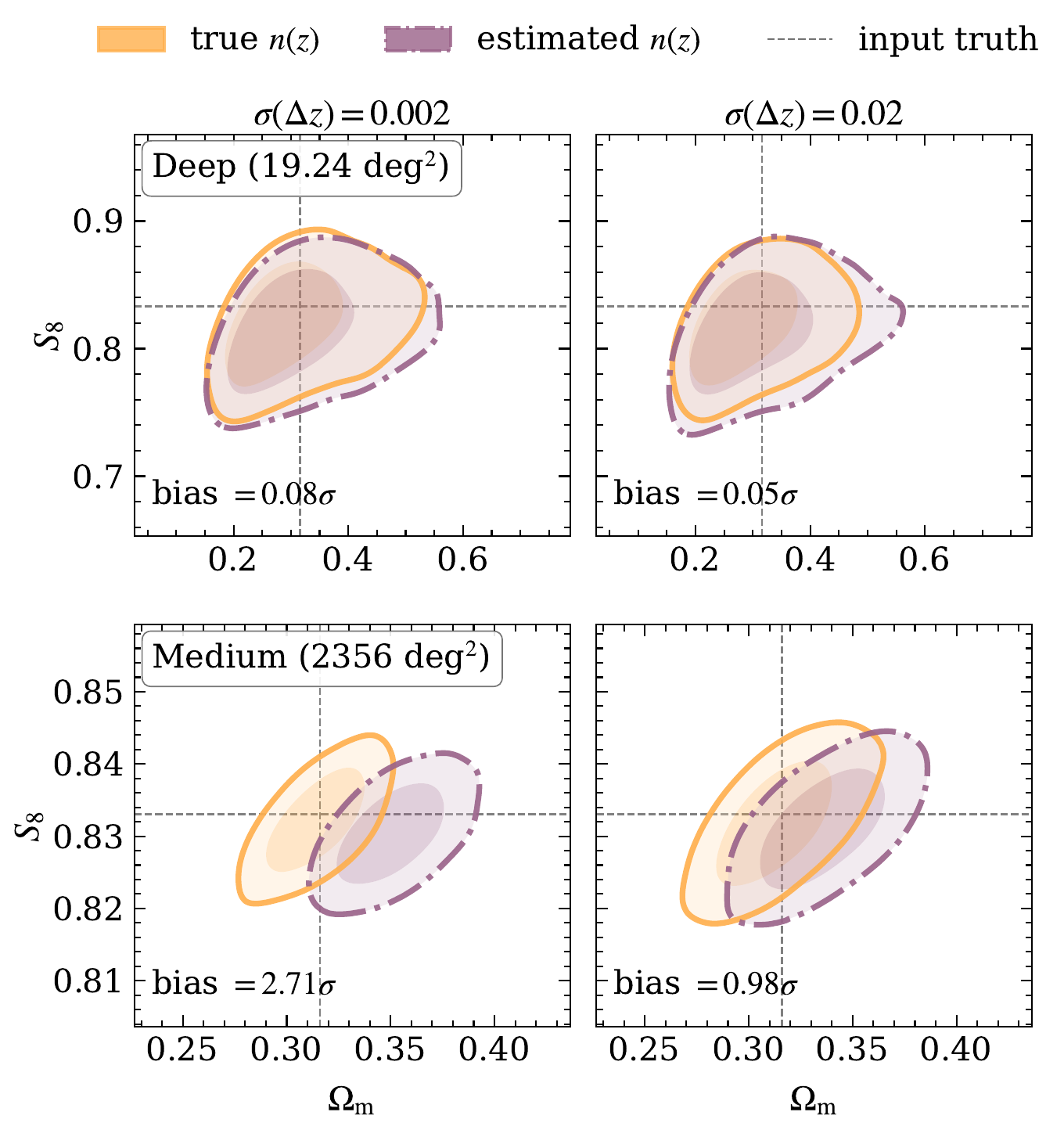}
    	\caption{Expected $\Omega_{\rm m}$–$S_8$ constraints from Roman cosmic shear analyses. Contours show the $68\%$ and $95\%$ credible regions obtained using the true redshift distributions (yellow) or the Roman-SOMPZ estimates (purple). Gray dashed lines mark the input cosmology used to generate the data vectors. The left and right columns assume $\sigma(\Delta_z)$ priors of $0.002$ and $0.02$, respectively. The top and bottom rows correspond to the Roman deep and medium tiers, respectively. }  
        \label{fig:nz_cosmology}
    \end{figure}

While the $n(z)$ comparisons in Fig.~\ref{fig:Nz} and Table~\ref{tab:mean_redshift} are promising, it is important to quantify how residual redshift biases propagate to Roman cosmic shear constraints. We therefore perform simulated likelihood analyses, generating theoretical data vectors and a covariance matrix using the true redshift distributions and analyzing them with the true $n(z)$ and the Roman-SOMPZ estimates to compute the model predictions. We adopt the galaxy number density, shear-calibration priors, and intrinsic-alignment priors of \citet{RomanDC1}, with a scale cut of $\ell_{\rm max}=3000$. For the cosmological parameters, we adopt the same priors as the Dark Energy Survey Year 3 (DES Y3) analyses \citep{desy3}.  We consider both the fiducial and conservative photometric-redshift priors of $\sigma(\Delta z)=0.002$ and $0.02$ \citep{Eifler21}, respectively. Here, $\Delta z$ is the mean-redshift bias in each tomographic bin, and both Gaussian priors are centered on zero.

Figure~\ref{fig:nz_cosmology} shows the resulting $\Lambda$CDM constraints in the $S_8$--$\Omega_{\rm m}$ plane. The top row represents an early Roman deep-tier analysis, 
while the bottom row represents the full medium-tier analysis. For the deep tier, the shifts in the $S_8$--$\Omega_{\rm m}$ plane are $0.08\sigma$ and $0.05\sigma$ under the fiducial and conservative redshift priors, respectively, indicating that Roman-SOMPZ is sufficiently accurate for early Roman weak lensing cosmic shear analyses. For the medium tier, however, the shifts increase to $2.71\sigma$ and $0.98\sigma$, respectively, suggesting that further improvements in redshift characterization will be required. 

For comparison, Appendix~\ref{app:nzcosmo} presents the cosmological constraints obtained using DES-SOMPZ. Although both methods produce acceptably small cosmological biases for the deep tier, Roman-SOMPZ yields $1\sigma$ parameter uncertainties up to $20\%$ smaller. Roman-SOMPZ also substantially reduces the medium-tier cosmological biases. The proposed SuPR Deep Survey could help address the remaining biases by improving spectroscopic coverage of the Roman galaxy samples.

\section{Conclusion}
\label{sec:conclusion}

The High Latitude Imaging Survey (HLIS) of the Nancy Grace Roman Space Telescope will observe weak lensing galaxy samples deeper than those of current Stage III surveys, placing stringent requirements on the characterization of their redshift distributions \citep{photoz-review}. At such depths, spectroscopy available for redshift calibration will inevitably be incomplete for faint, high-redshift populations. In this work, we use simulated LSST+Roman multiband photometry from the \texttt{Cardinal} mock catalogs  (Fig.~\ref{fig:footprint}; \citealt{Cardinal}) to quantify the effect of this incompleteness on self-organizing map (SOM)-based redshift characterization. Using the DES Self-Organizing Map $p(z)$ (SOMPZ) framework \citep{SOM-DES_Y6, SOMPZ-DES_Y6} as a baseline, we develop Roman-SOMPZ to algorithmically mitigate spectroscopic selection bias without adding spectroscopic follow-up to the reference sample.

Our main findings are:
\begin{enumerate}

\item In the Roman-like \texttt{Cardinal} mocks, 
spectroscopic incompleteness causes DES-SOMPZ to exclude $7.2\%$ of wide-field galaxies from tomographic binning and leaves $26.6\%$ of deep-field galaxies without direct $p(z|c,\hat{b})$ calibration (Fig.~\ref{fig:SOMs}). 
The incomplete reference sample produces large overlaps between tomographic bins and biased $n(z)$ estimates at high redshift (Fig.~\ref{fig:Nz}, top-right). 

\item We address tomographic binning separately from redshift estimation. For binning, we use \texttt{FlexZBoost} point estimates for deep-field galaxies to assign wide SOM cells and Jenks natural breaks to optimize bin edges. This recovers galaxies in previously unassigned wide-SOM cells and reduces overlaps between tomographic bins (Fig.~\ref{fig:bin_overlaps}). 

\item For redshift calibration, we introduce UMAP-flow-$z$, a UMAP+normalizing flow model that estimates redshifts for galaxies in no-spec deep-SOM cells. Augmenting the redshift-calibration sample using these estimates reduces the residual high-redshift $n(z)$ biases (Fig.~\ref{fig:Nz}, bottom-right).

\item Roman-SOMPZ reduces the absolute mean-redshift biases in eight of nine bins relative to DES-SOMPZ (Table~\ref{tab:mean_redshift}), bringing the recovered $n(z)$ closer to those obtained with representative spectroscopy (Fig.~\ref{fig:Nz}, top-left vs. bottom-right). Bins~4--6 reach the Roman target of $|\Delta\langle z\rangle|/(1+\langle z\rangle)<0.002$ \citep{RomanSRD}, while larger biases remain at lower and higher redshifts.

\item In the simulated cosmic shear analyses, we find that Roman-SOMPZ is sufficiently accurate for the early deep-tier cosmological constraints. Full medium-tier cosmology will require further improved redshift characterization, likely enabled by deeper spectroscopy such as the proposed Subaru-PFS/Roman (SuPR) Deep Survey (Fig.~\ref{fig:nz_cosmology}). 

\end{enumerate}

These results support the use of Roman-SOMPZ for early Roman weak lensing analyses before representative spectroscopic samples become available. Although we develop the method for Roman HLIS, the approach can be applied to similar imaging surveys that face incomplete spectroscopy. These include Rubin LSST and the Euclid Wide Survey, where shallower source samples are expected to make this problem less challenging than for Roman. 

Our tests use a single simulation. 
Going forward, we will examine robustness across galaxy--halo connection models such as \texttt{Diffsky} \citep{Diffstar, DSPS} and use those tests to quantify the augmentation uncertainty. Future work can also explore combining Roman-SOMPZ with clustering-redshift measurements to optimize $n(z)$ characterization. By showing that task-specific use of individual-galaxy redshift estimators can maximally recover redshift information lost to spectroscopic incompleteness, this work offers a scalable route to robust photometric-redshift characterization for Roman and other Stage~IV surveys. 

\textit{Data availability.} The \texttt{Roman-SOMPZ} code, along with the configurations used in this work, will be made publicly available upon acceptance of this paper.

\section*{Acknowledgments}
YZ is supported with funding by the University of Chicago Data Science Institute (DSI). CHT is supported by the Eric and Wendy Schmidt Fellowship. This work was partially supported by NASA grant 22-ROMAN22-0011 Maximizing Cosmological Science with the Roman High Latitude Imaging Survey Project Infrastructure Team (PIT). This work was completed in part with resources provided by the University of Chicago’s Research Computing Center.

\bibliographystyle{aasjournal}
\bibliography{oja_template}
\begin{appendix}

\section{UMAP-flow-$z$ methodology and implementation}
\label{app:umap-flow-z}

This appendix describes the UMAP-flow-$z$ estimator used to augment the redshift-calibration sample in Section~\ref{sec:umap}. 
We first generate a continuous low-dimensional representation of the deep-field photometry using the UMAP approach and code of \citet{UMAP-kNN-z}, then train a normalizing flow to learn the joint distribution of the three UMAP coordinates and spectroscopic redshift. At each target galaxy's UMAP coordinates, we evaluate and normalize the joint density as a function of redshift to obtain a conditional distribution. We adopt the median of this distribution as the galaxy's redshift point estimate. For galaxies in no-spec deep-SOM cells, these point estimates augment the redshift-calibration sample, which then enters 
the standard SOMPZ procedure for estimating the ensemble $n(z)$ distributions. 

\subsection{UMAP embedding and reference sample}
\label{app:umap-flow-z-inputs}

Following \citet{UMAP-kNN-z}, we construct the three-dimensional embedding from seven deep-field colors: 
\begin{equation}
    \boldsymbol{x}
    = (u-g,\,g-r,\,r-i,\,i-z,\,z-y,\,y-J,\,J-H). 
\end{equation}
UMAP \citep{UMAP} maps these features to coordinates $\boldsymbol{u}=(u_1,u_2,u_3)$.
The UMAP embedding is fitted on the full deep-field population, placing both the spec-$z$ reference galaxies and target galaxies in a common coordinate system.

The spectroscopic reference sample contains the 28,583 COSMOS-matched spec-$z$ galaxies described in Section~\ref{sec:data}, spanning approximately $0.019<z<2.283$. The remaining 3,417,426 deep-field galaxies constitute the test sample. 
Figure~\ref{fig:UMAP} provides a complementary view of the three samples compared in Fig.~\ref{fig:SOM-augmented}, showing galaxy-overlaid three-dimensional UMAP embedding rather than cell-averaged redshifts on the deep SOM. From top to bottom, the rows show the COSMOS-matched spec-$z$, the reference sample augmented with UMAP-flow-$z$ estimates, and the deep-field population colored by simulation-truth redshifts. The embedding is shown from four viewing directions to display the three-dimensional geometry. Section~\ref{app:umap-flow-z-prediction} describes the augmented redshift-reference sample (middle row) in detail.  

The spec-$z$ galaxies in the top row of Fig.~\ref{fig:UMAP} trace a thin manifold with a smooth, approximately monotonic redshift trend and no apparent gaps across. The sampling is non-uniform and becomes sparser toward high redshift. The deep-field photometric sample shown in the bottom row populates a manifold that extends slightly beyond the region covered by spec-$z$, 
in a direction approximately orthogonal to the main redshift gradient. 
The well-behaved embedding and smooth redshift trend motivate using the spec-$z$ to infer redshifts at the UMAP coordinates of galaxies in no-spec deep-SOM cells. 

\subsection{Joint normalizing flow model and training}
\label{app:umap-flow-z-model}

Before fitting the density model, we standardize each UMAP coordinate and the spectroscopic redshift:
\begin{equation}
    \widetilde{u}_i = \frac{u_i-\overline{u}_i}{s_{u_i}},
    \qquad
    \widetilde{z} = \frac{z-\overline{z}}{s_z},
    \label{eq:umap-flow-z-standardization}
\end{equation}
where $i=1,2,3$. The means and population standard deviations are computed over the full spectroscopic reference sample before the internal training--validation split. For this sample, $\overline{z}\simeq0.756$ and $s_z\simeq0.412$. We apply the same standardization factors to the non-spectroscopic galaxies at prediction time. The flow input is 
\begin{equation}
    \boldsymbol{v}
    = (\widetilde{u}_1,\widetilde{u}_2,\widetilde{u}_3,
       \widetilde{z}).
\end{equation}

We model the joint density $q_{\theta}(\boldsymbol{v})$ with a masked autoregressive flow \citep{flow-MAF} using rational-quadratic spline transformations \citep{flow-transformer}, implemented in \texttt{FlowJAX}.\footnote{\url{https://danielward27.github.io/flowjax/}} This is an unconditional four-dimensional density estimator, and we derive the conditional redshift distributions after training. 

Let $T_{\theta}$ denote the invertible transformation from standardized data coordinates $\boldsymbol{v}$ to latent coordinates. The model density is
\begin{equation}
    q_{\theta}(\boldsymbol{v})
    = \mathcal{N}_4\!\left(T_{\theta}(\boldsymbol{v});
       \boldsymbol{0},\mathbf{I}\right)
      \left|\det\frac{\partial T_{\theta}}
            {\partial\boldsymbol{v}}\right|,
    \label{eq:umap-flow-z-density}
\end{equation}
where the base distribution is a four-dimensional standard normal. Each autoregressive transformation uses a masked neural network to parameterize scalar spline transformations, with coordinate permutations between successive flow layers. We construct the model using \texttt{masked\_autoregressive\_flow} and a \texttt{RationalQuadraticSpline} transformer with \texttt{knots=12} and \texttt{interval=5}. The latter sets the spline interval to $[-5,5]$; it does not impose a bound on the physical redshifts. For the remaining architectural settings, we retain the defaults in \texttt{FlowJAX} version~19.0.0, using eight autoregressive flow layers. Each layer uses a masked neural network with one hidden layer of 50 units and rectified linear unit (ReLU) activation.

We train the model by minimizing the unweighted mean negative log likelihood, 
\begin{equation}
    \mathcal{L}(\theta)
    = -\frac{1}{N_{\rm fit}}
       \sum_{j\in\mathcal{S}_{\rm fit}}
       \ln q_{\theta}(\boldsymbol{v}_j), 
    \label{eq:umap-flow-z-loss}
\end{equation}
where $\mathcal{S}_{\rm fit}$ is the fitting subset of the spectroscopic reference sample and $N_{\rm fit}=|\mathcal{S}_{\rm fit }|$. 
We use the default \texttt{MaximumLikelihoodLoss} in \texttt{FlowJAX}. The \texttt{fit\_to\_data} function randomly reserves 10\% of the spectroscopic sample for validation and uses the remaining 90\% for fitting. We set the learning rate to $3\times10^{-4}$, the batch size to 512, the maximum training to 500 epochs, and an early-stopping patience to 30 consecutive epochs without improvement in the validation loss. The model parameters with the lowest validation loss are retained.

\subsection{Conditional redshift distributions and point estimates}
\label{app:umap-flow-z-prediction}

For a target galaxy at UMAP coordinates $\boldsymbol{u}_j$, we evaluate the trained joint density as a function of redshift. Normalizing this one-dimensional slice gives the conditional density over the adopted evaluation interval:
\begin{equation}
    \widehat{p}_{\theta}(z\mid\boldsymbol{u}_j)
    =
    \frac{
       q_{\theta}\!\left(
       \widetilde{\boldsymbol{u}}_j,
       (z-\overline{z})/s_z\right)
    }{
       \displaystyle\int_{0}^{3}
       q_{\theta}\!\left(
       \widetilde{\boldsymbol{u}}_j,
       (z'-\overline{z})/s_z\right)\,\mathrm{d}z'
    },
    \qquad 0\leq z\leq3.
    \label{eq:umap-flow-z-conditional}
\end{equation}
The constant Jacobian factors from standardization cancel between the numerator and denominator. This step conditions the fitted joint distribution on the galaxy's position in UMAP space. 

In practice, we evaluate the density at 151 uniformly spaced redshifts, $z_k=0.02k$ for $k=0,\ldots,150$. The grid is transformed to standardized redshift before evaluating the flow. We define 
\begin{equation}
    \ell_{jk}
    = \ln q_{\theta}\!\left(
      \widetilde{\boldsymbol{u}}_j,
      (z_k-\overline{z})/s_z\right).
\end{equation}
To avoid numerical underflow, we subtract the maximum log density along each galaxy's redshift grid before exponentiation. The normalized discrete weights are
\begin{equation}
    w_{jk}
    = \frac{\exp(\ell_{jk}-\ell_{j,\max})}
           {\sum_{m=0}^{150}\exp(\ell_{jm}-\ell_{j,\max})},
    \qquad
    \ell_{j,\max}=\max_k\ell_{jk}.
    \label{eq:umap-flow-z-weights}
\end{equation}
These weights satisfy $\sum_k w_{jk}=1$ and provide a rectangular approximation to Eq.~\ref{eq:umap-flow-z-conditional}. The corresponding density values in physical redshift units are $\widehat{p}_{jk}=w_{jk}/\Delta z$, where $\Delta z=0.02$.

\begin{figure*}[!t]
    \centering
    \includegraphics[width=\linewidth]{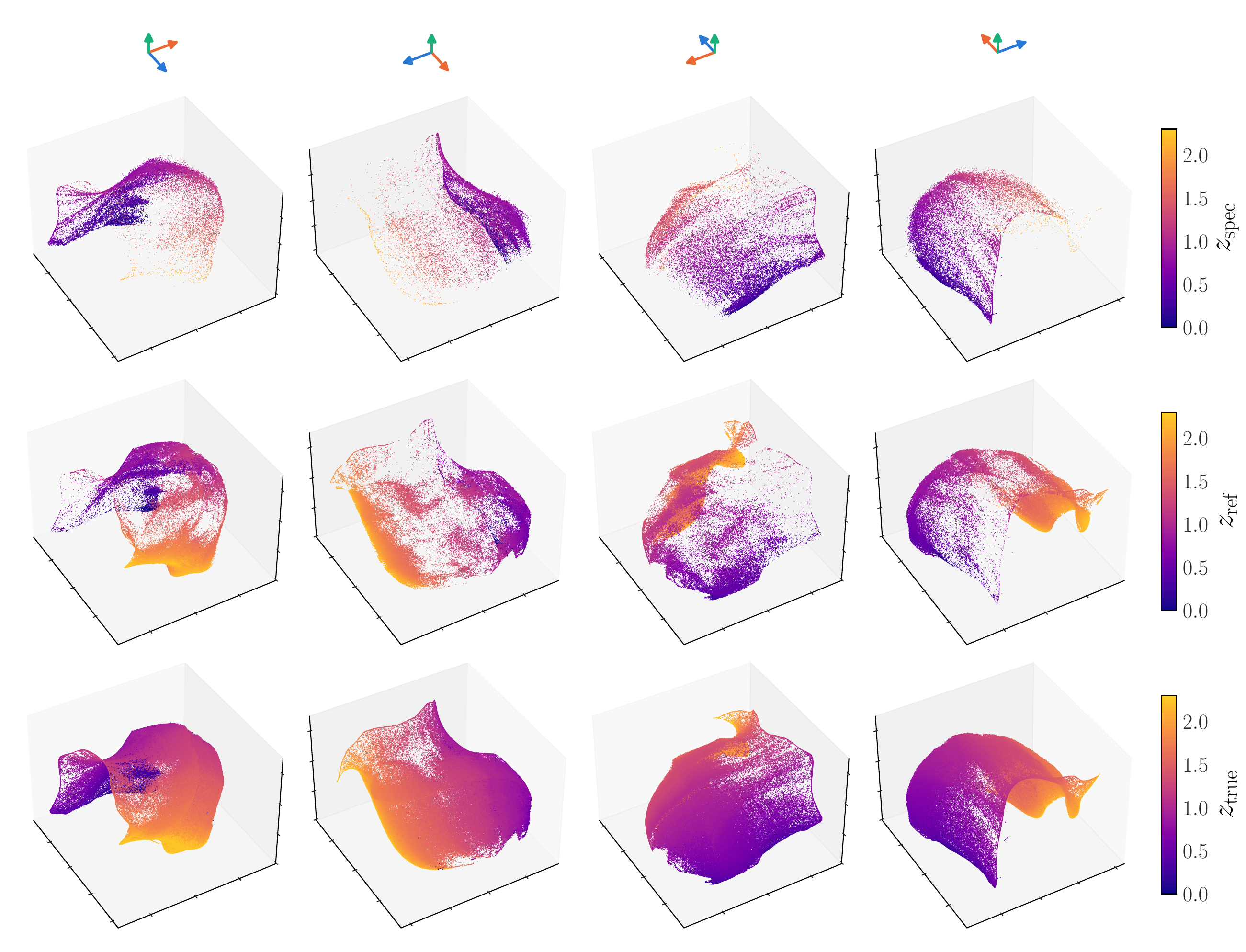}
    \caption{Three-dimensional UMAP embedding used by the UMAP-flow-$z$ estimator, constructed from seven deep-field colors: $u-g,\,g-r,\,r-i,\,i-z,\,z-y,\,y-J,\,J-H$. From top to bottom, the three rows correspond to samples in the left, middle, and right panels of Fig.~\ref{fig:SOM-augmented}, shown here in the UMAP space from four viewing directions. Each column uses the same viewing direction for all three rows, with successive columns rotated by $90^\circ$ about the vertical UMAP axis. The arrows above each column indicate the projected directions of the three UMAP axes. Numerical axis values are omitted because UMAP coordinates have no physical meaning. Each point represents a \texttt{Cardinal} deep-field galaxy at its UMAP coordinates. 
    \textit{Top row:} The COSMOS-matched spec-$z$ sample used for normalizing flow training, colored by high-confidence redshifts. 
    \textit{Middle row:} Approximately 10\% of the augmented redshift-reference sample used for $n(z)$ estimation in Section~\ref{sec:umap}. The 94,683 plotted galaxies are colored by their reference redshifts: high-confidence spec-$z$ for 2,811 galaxies and UMAP-flow-$z$ estimates for 91,872 galaxies in no-spec deep-SOM cells.
    \textit{Bottom row:} A 10\% subsample of the full deep-field photometric sample, colored by true redshifts. This subsample includes every galaxy shown in the middle row.
    The spec-$z$ galaxies trace a thin manifold with redshift varying smoothly and monotonically across it. The sampling is non-uniform and becomes sparser toward high redshift but shows no apparent gaps. Near the manifold edge, the photometric sample shown in the bottom row extends slightly beyond the region covered by spec-$z$, in a direction approximately orthogonal to the redshift variation. 
    }
    \label{fig:UMAP}
\end{figure*}

Finally, we approximate the posterior median on this grid to obtain the point estimates used in Section~\ref{sec:umap}. 
The evaluation interval extends beyond the \texttt{Cardinal} cutoff at $z\simeq2.3$, and we do not clip predictions at that cutoff. 
We generate predictions for all non-spectroscopic deep-field galaxies but add only those in no-spec deep-SOM cells to the redshift-reference sample, and the spec-$z$'s are retained. 
The middle row of Fig.~\ref{fig:UMAP} shows a subsample of this augmented reference catalog, colored by spectroscopic redshift or the UMAP-flow-$z$ point estimate. Compared with the top row, the augmented sample enriches the reference redshifts in poorly sampled regions of the embedding, including high-redshift regions. 

The middle and bottom rows of Fig.~\ref{fig:UMAP} show that the augmented reference redshifts reproduce the true-redshift trends of the deep-field population along the UMAP manifold. This qualitative check of the inferred redshifts in UMAP space complements the cell-averaged comparison in Fig.~\ref{fig:SOM-augmented} and the individual-galaxy assessment in Fig.~\ref{fig:UMAP-flow-z}. The augmented reference sample is then propagated through Eq.~\ref{eq:sompz} of SOMPZ to estimate $n(z)$ for each tomographic bin.

\section{Impacts of different $n(z)$ estimations on cosmological constraints} 
\label{app:nzcosmo}

\begin{figure*}[!t]
        \centering
\includegraphics[width=0.5\linewidth]{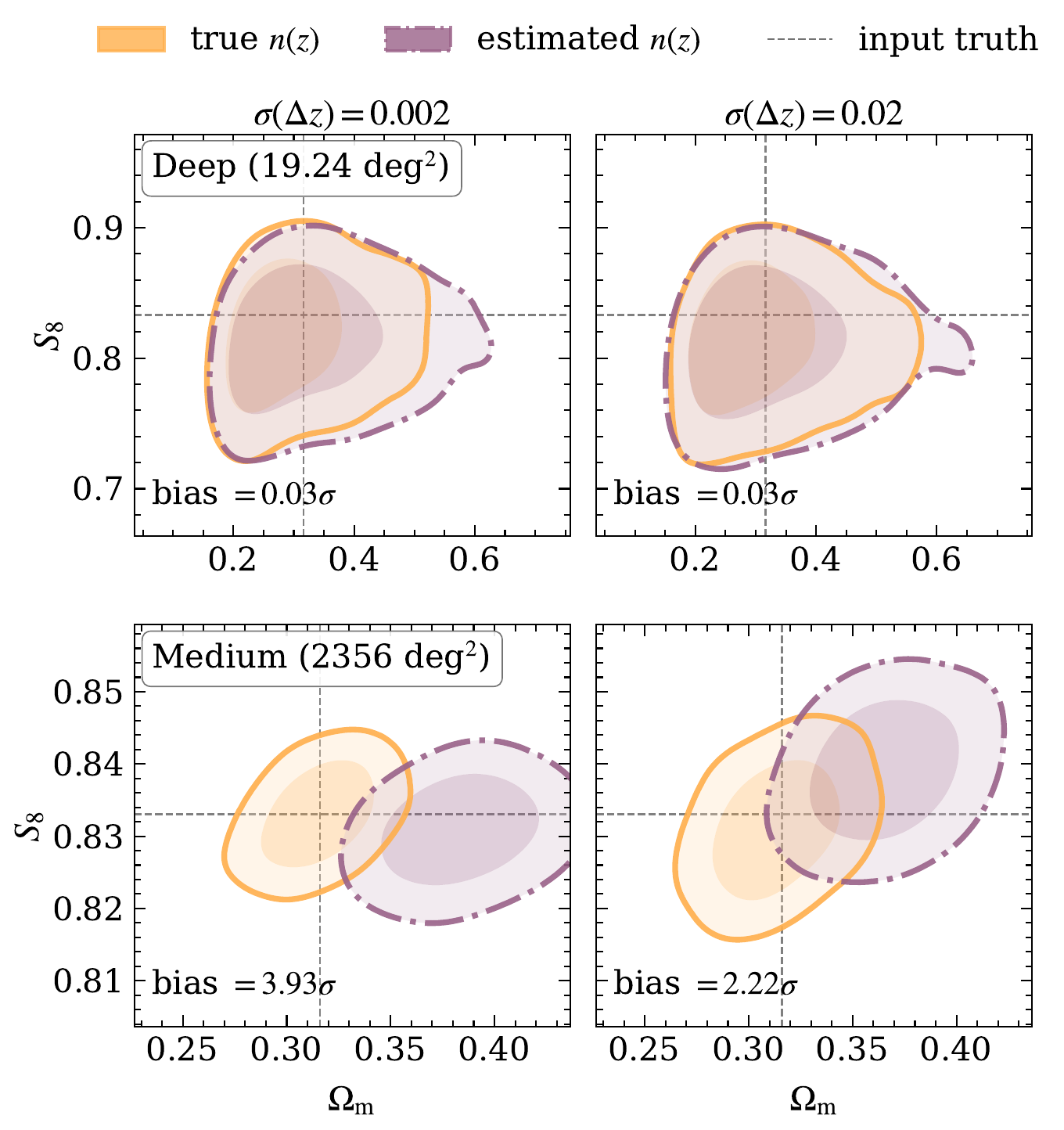}
    	\caption{Same as Fig.~\ref{fig:nz_cosmology}, but using the true and estimated $n(z)$s based on the DES-SOMPZ algorithm.}  
        \label{fig:nz_cosmology_des}
    \end{figure*}

In Section~\ref{sec:cosmo}, we present the expected cosmic shear cosmological constraints obtained using Roman-SOMPZ. Here, we repeat the analysis with DES-SOMPZ to assess how the methodological improvements introduced in Roman-SOMPZ affect cosmological constraints. Figure~\ref{fig:nz_cosmology_des} shows the resulting $\Lambda$CDM constraints in the $S_8$--$\Omega_{\rm m}$ plane, using the same analysis settings as Fig.~\ref{fig:nz_cosmology}.

For the deep tier, DES-SOMPZ also produces acceptably small cosmological biases, but weaker constraints than Roman-SOMPZ. The $1\sigma$ uncertainties on $S_8$ increase by $25\%$ and $22\%$ 
under the priors of $\sigma(\Delta_z)=0.02$ and $0.002$, respectively. The corresponding increases for $\Omega_{\rm m}$ are $20\%$ and $15\%$. For the medium tier, replacing Roman-SOMPZ with DES-SOMPZ increases the bias on $\Omega_{\rm m}$--$S_8$ plane from $0.98\sigma$ to $2.22\sigma$ for $\sigma(\Delta_z)=0.02$, and from $2.71\sigma$ to $3.93\sigma$ for $\sigma(\Delta_z)=0.002$. These results demonstrate the impact of redshift characterization method on both the precision and accuracy of cosmological constraints.

\end{appendix}

\end{document}